\documentstyle[epsf,aps,multicol,emlines]{revtex}
\draft
\author{{R.~Folk$^{(1)}$, Yu.~Holovatch$^{(2,3)}$, T.~Yavors'kii$^{(3)}$}
\\[1.5ex]
$^{(1)}$Institut f\"ur theoretische Physik,
Johannes Kepler Universit\"at  Linz,
A--4040 Linz, Austria\\
E-mail: folk@tphys.uni-linz.ac.at
\\
$^{(2)}$Institute for Condensed Matter Physics, National Academy
of Sciences of Ukraine, UA--79011 Lviv, Ukraine\\ E-mail:
hol@icmp.lviv.ua
\\
$^{(3)}$Ivan Franko National University of Lviv, UA--79005 Lviv, Ukraine\\
E-mail: tarasyk@ktf.franko.lviv.ua
}

\title{
Critical exponents of a three dimensional weakly diluted quenched
Ising model}

\date{\today}
\begin{document}
\maketitle
\begin{abstract}
{\small We discuss universal and non--universal critical exponents
of a three dimensional Ising system in the presence of weak
quenched disorder. Both experimental,
computational, and
theoretical results are reviewed. Special attention is paid to the
results obtained by the field theoretical renormalization group
approach. Different renormalization schemes are considered putting
emphasis on analysis of divergent series obtained. }
\end{abstract}




\section{Introduction}

\label{I} 
\begin{figure} [htbp]
\epsfxsize 50mm \centerline{\epsffile{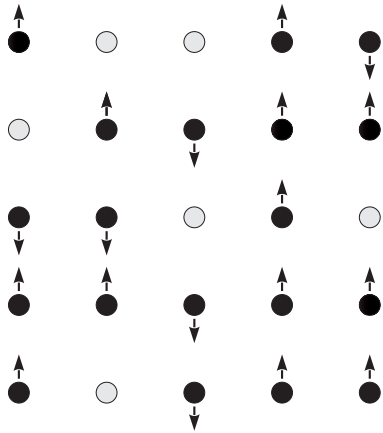}}
\vspace*{5mm}
\caption
{Weakly diluted quenched 3d Ising model (random Ising model: RIM)
describing a system of scalar "spins" randomly distributed in
sites of a three-dimensional cubic lattice and fixed in certain
positions.\label{figlattice} }
\end{figure}

The present paper reviews properties of a
three-dimensional weakly diluted quenched Ising model (random
Ising model: RIM -- see figure \ref{figlattice}) in the vicinity
of critical point. For weak dilution by a non-magnetic component,
i.e. far from the percolation threshold, the magnetic second-order
phase transition still maintains in the RIM \cite{random} although it
possesses novel features in comparison with the pure $d=3$ Ising
model. Static critical exponents of a RIM \cite{note} have been a
subject of a detailed
experimental \cite{Dunlap81,Birgeneau83,Hastings85,Belanger86,Barret86,%
Mitchell86,Thurston88,Rosov88,Belanger88,Ramos88,Ferreira91,%
Belanger95,Belanger96,Hill97,Slanic98a,Slanic98,Slanic99},
numerical \cite{Landau80,Marro86,Chowdhury86,Braun88,Braun89,Wang89,Wang90,%
Holey90,Heuer90,Heuer93,Prudnikov93,Hennecke93,Parisi98,Wiseman98a,%
Wiseman98b,Aharony98,Marques00,Marques00a,Marques00b,Hukushima00}
and theoretical
\cite{Harris74b,Khmelnitskii75,Lubensky75,Grinstein76,Jayaprakash77,%
Shalaev77,Sokolov77,Sokolov81,Newman82,Jug83,Mayer84,Shpot89,Mayer89,%
Mayer89a,Holovatch92,Janssen95,Shalaev97,Folk98,Holovatch98,Folk99,%
Varnashev00,Folk00,Pakhnin00,Pelissetto00}
analysis almost for three decades. Recently, new data were
obtained both in experimental measurements
\cite{Belanger95,Belanger96,Hill97,Slanic98a,Slanic98,Slanic99}
and Monte Carlo simulations
\cite{Parisi98,Wiseman98a,Wiseman98b,Aharony98,Marques00,Marques00a,%
Marques00b,Hukushima00}. Theoretical breakthrough occurred along
several months in 1999--2000 when the perturbation expansion series
for RIM were extended from the 4th \cite{Mayer89} through 5th
\cite{Folk00,Pakhnin00} to the sixth order
\cite{Carmona00,Pelissetto00}. Therefore it seems for us worth collecting
together the bulk of available results concerning critical
behaviour of the RIM.

In the present discussion, we will be mainly interested in the RIM
critical exponents. An {\em asymptotic critical exponent} $x$ of a
physical observable ${\cal O}(\tau)$ is defined \cite{Stanley71}
asymptotically close to the critical point $T_c$:
$x \equiv \lim_{\tau\rightarrow 0} \frac{\ln {\cal
O}(\tau)}{\ln|\tau|}$, where $\tau = (T-T_c)/T_c$ is the reduced
distance to the critical point. For instance, the magnetic
susceptibility $\chi$ diverges as:
\begin{equation} \label{1}
\chi \simeq \Gamma_{\pm}|\tau|^{-\gamma}\,, \mbox{\hspace{2em}}\tau \rightarrow
0
\end{equation}
where
$\gamma$ is the susceptibility critical exponent, while $\Gamma_{+}$ and
$\Gamma_{-}$ are critical amplitudes above and below the critical point
respectively.
The power law of type (\ref{1}) holds exactly in the asymptotic
 regime $\tau \rightarrow 0$. In this regime the critical exponents and ratios
of the critical amplitudes take constant values. According to the
universality hypothesis they are defined by global variables only.
For a short-range interaction the global variables are the space
dimension and tensor characteristics of an order parameter. In the
non-asymptotic regime the approach to criticality is described by
non-universal {\em effective critical exponents}, which are
introduced to describe the behaviour of a quantity in a certain
temperature interval \cite{Kouvel64,Riedel74}. The susceptibility
effective critical exponent $\gamma_{_{eff}}$ by definition reads:
\begin{equation} \label{2}
\gamma_{_{eff}}(\tau) = \frac{{\rm d} \ln\chi(\tau)}{{\rm d} \ln\tau}.
\end{equation}
In the asymptotic limit $\tau\rightarrow 0$ the effective and asymptotic
exponents coincide. In the intermediate region the behaviour is characterized
by the so--called Wegner expansion \cite{Wegner72}:
\begin{equation}
\label{3} \chi \simeq \Gamma_{\pm}
|\tau|^{-\gamma}\left(1+\Gamma_{1,\pm}|\tau|^{\omega\nu}
+\Gamma_{2,\pm}|\tau|^{2\omega\nu}+\ldots\right)\, ,
\end{equation}
where $\Gamma_{1,\pm},\Gamma_{2,\pm}$ are non--universal
amplitudes, $\nu$ is the correlation length critical exponent,
and $\omega$ is the correction--to--scaling exponent.

It is the art of a physicist which allows to discriminate between
regimes (\ref{1})--(\ref{3}) either performing experimental
measurements or doing theoretical calculations and numerical
simulations. Below we will review available data on effective and
asymptotic critical exponents of the RIM. The paper is arranged as
follows. In the next section \ref{II} we will give general
considerations about an influence of a weak quenched disorder on a
second order phase transition, formulate the model as well as
outline the main features of its critical behaviour. In
Section \ref{III} we review experiments on weakly diluted uniaxial
magnets, whereas computer Monte Carlo simulations of a RIM are
analyzed in Section \ref{IV}. In sections \ref{V} and \ref{VI} we
outline the method of renormalization group as the most fruitful
theoretical tool in the study of the critical properties of RIM.
We consider different renormalization schemes and pay special
attention to analysis of divergent series obtained within the
approach. Conclusions (Section \ref{VII}) complete the article.


\section{Weakly diluted quenched Ising model}
\label{II}

The central questions one has to answer studying the influence of
weak disorder on a magnetic second-order phase transitions are: do
the critical exponents of a homogeneous magnet change under
dilution by a non-magnetic component? And if this is the case,
are the new exponents universal? Regarding the first question it was
argued \cite{Harris74} that if the heat capacity critical exponent
$\alpha$ of the pure (undiluted) system is positive, i.e. the heat
capacity diverges at the critical point, then a quenched disorder
causes changes in the critical exponents. This statement is known
as Harris criterion. Later, for a large class of
$d$--dimensional disordered systems it was proven that the
correlation length critical exponent $\nu$ must satisfy  the bound
$\nu \geq 2/d$ \cite{Chayes86}. Both statements focus attention
towards studies of the $d=3$ Ising model, where the typical
numerical values of the above exponents, together with the
magnetic susceptibility and the order parameter critical exponents
in the pure case read \cite{note1}:
\begin{equation}  \label{4}
\alpha=0.109 \pm 0.004 >0, \, \nu=0.6304 \pm 0.0013<2/3, \,
\gamma=1.2396 \pm 0.0013, \, \beta= 0.3258 \pm 0.0014.
\end{equation}
According the inequalities of the criteria mentioned above
for the case of the diluted Ising model new exponents are expected.

In order to obtain precise values of critical exponents it is
standard now to rely on renormalization group methods. In
particular, the theoretical estimates given above (\ref{4}) were
obtained on the basis of a deep analogy between long-distance
properties of the Ising model in the neighbourhood of a
second-order phase transition point and a field theory with an
effective Landau-Ginzburg-Wilson Hamiltonian of the form:
\begin{equation} \label{5}
{\cal H}_{\rm Ising}(\varphi)=\int {\rm d}^3R \Big\{ {1\over 2}
\left[|\nabla \varphi|^2+ m_0^2 {\varphi}^2\right] + {\tilde
u_0\over 4!} \varphi^4\Big\},
\end{equation}
where $m_0^2$ is a bare mass proportional to distance to a critical point,
$\varphi=\varphi(R)$, $\tilde u_0$ are (bare) scalar field and coupling.

One of the ways to introduce quenched dilution to the effective
Hamiltonian (\ref{5}) is to add to $m_0^2$ the random-temperature
like variable \cite{Grinstein76}  $\psi=\psi(R)$:
\begin{equation} \label{6}
{\cal H}_{\psi}(\varphi)=\int {\rm d}^3R \Big\{ {1\over 2}
\left[|\nabla \varphi|^2+ (m_0^2+\psi) \varphi^2\right] + {\tilde
u_0\over 4!} \varphi^4\Big\}.
\end{equation}
Taking that  $\psi$ obeys Gaussian distribution:
\begin{equation} \label{7}
P(\psi)=\frac{1}{\sqrt{4\pi}w}\exp(-\psi^2/4w^2)
\end{equation}
with $w^2$ being a half of dispersion,
and introducing $n$ replicas \cite{replicas}
of a model (\ref{5}) in order to perform averaging over
quenched disorder \cite{Brout59}
one ends up \cite{Grinstein76} with a familiar effective Hamiltonian:
\begin{equation} \label{8}
{\cal H}_{\rm RIM}(\varphi)=
\int {\rm d}^3R \Big\{ {1\over 2} \sum_{\alpha=1}^{n}
\left[|\nabla {\varphi}_\alpha|^2+ m_0^2 {\varphi}_\alpha^2\right] +
{u_{0}\over 4!}
\sum_{\alpha=1}^{n}{\varphi}_\alpha^4 +
{v_{0}\over 4!} \left(\sum_{\alpha=1}^{n}{\varphi}_\alpha^2 \right)^2
\Big\}.
\end{equation}
In the limit  $n\rightarrow 0$ field theory (\ref{8}) describes
critical properties of a RIM. Here, the bare coupling $u_0$ is
positive (being proportional to $\tilde u_0$) whereas the bare
coupling $v_0$ is proportional to minus variance of the random
variable $\psi$ and thus is negative. The last term in (\ref{8})
is present only for non-zero dilution: it is  directly responsible
for the fluctuations effective interaction due to the presence of
impurities.

Often (e.g. in Monte Carlo simulations) the microscopic
Hamiltonian of RIM is written in the following form:
\begin{equation} \label{9}
H =  - \frac{1}{2}\sum_{R, R^{\prime}}
J(|{R - R^{\prime}}|) {S}_{R} {S}_{R^{\prime}}
c_{R} c_{R^{\prime}},
\end{equation}
where $R$ runs over sites of simple cubic lattice, $J$ is a short--range
translationally invariant interaction between pairs of $S=\pm 1$ ``classical''
Ising spins, $c_R$ is an occupation number equal either to 0 or to 1.
It is considered that geometrically the
vacancies $c_R=0$ are distributed independently  according to the law
\begin{equation} \label{10}
P(c_R)=(1-p)\delta(c_R)+p\delta(1-c_R)
\end{equation}
and fixed in different sites of the lattice (see figure
\ref{figlattice}). In (\ref{10}) $p\leq 1$ is the concentration of
occupied sites. Calculating free energy of a model (\ref{9}) and
using replica trick \cite{replicas} to perform the configurational
averaging of the logarithm of configuration-dependent partition
function  it is straightforward to show that one ends up with the
effective Hamiltonian of (\ref{8}) type in the $n \rightarrow 0$
replica limit with $u_0\sim p$ and $v_0\sim p(p-1)$.

From the viewpoint of dynamics one can point to two opposite types
of a disorder \cite{note}. If a characteristic time of impurities
dynamics is comparable to relaxation times in the pure system,
impurity variables are treated identically to the ``pure''
dynamical variables and are a part of the disordered system phase
space. The corresponding annealed disorder \cite{Brout59} is a
subject of special investigations and reveals trivial results at
criticality. The so called Fisher renormalization \cite{Fisher68}
states that when a heat capacity critical exponent of an undiluted
system $\alpha_{pure}$ is positive, then the critical exponents
$x$ of an annealed system are determined by those of the
corresponding pure one ($x_{pure}$) by a simple renormalization of
the form:
\begin{equation}\label{10a}
x = \frac{x_{_{pure}}}{1-\alpha_{_{pure}}}, \, \,
\alpha = \frac{-\alpha_{_{pure}}}{1-\alpha_{_{pure}}}.
\end{equation}
This explains why prevailing interest is attracted by the quenched disorder
when impurities can be considered as fixed and thus one needs to perform
configurational average over an ensemble of disordered systems with different
realization of the disorder.

The Hamiltonian (\ref{8}) represents critical properties of the
problem (\ref{9}) for small randomness.
Alternatively, the scale-invariant fractal ramified
cluster at the percolation threshold is a starting point of the strong
disorder approach. Here, the field theoretical description starts with
the effective Hamiltonian of the Potts model. A unified theory
of random systems critical behaviour which would give both  regimes
of strong disorder and weak disorder as its limiting cases is still absent.

The translation invariant Hamiltonian (\ref{8}) presumes perturbative
account of thermal fluctuations around a spatially homogeneous
unique ground
state. While it appears to hold for a pure system, for a random
system there is a macroscopic number of spatially inhomogeneous
ground states in a disorder dominated region. These correspond to
local minimum solutions of a saddle-point equation for the
effective Hamiltonian (\ref{8}) \cite{Dotsenko95,Dotsenko98}.
Physically the last corresponds to the so--called Griffiths phase
\cite{Griffiths69} caused by the existence of ferromagnetically
ordered ``islands'' in the temperature interval between the
critical temperatures of pure and random systems. The description
of the phase at the critical point is provided by a
replica--symmetry breaking Hamiltonian \cite{Dotsenko95} leading
to non-trivial results \cite{Dotsenko95a,Wu98}.
However recently a refined analysis of the problem brought about a
stability of the critical behaviour of the weakly disordered systems
with respect to replica symmetry breaking effects \cite{Prudnikov01}.
The theoretical
results reviewed in the present discussion will be based
on the replica symmetrical Hamiltonian (\ref{8}) (see section
\ref{V}).

\begin{figure} [htbp]
\epsfxsize 100mm \centerline{\epsffile{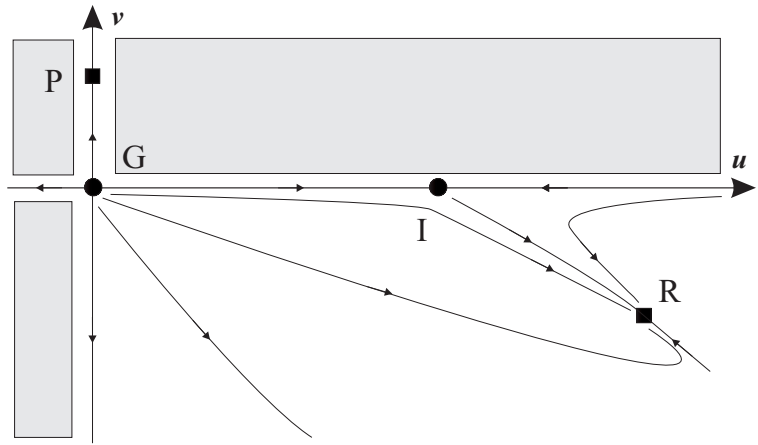}}
\vspace*{5mm}
\caption{RIM
fixed points qualitative structure. Gaussian fixed point {\bf G}
is stable for $d\geq4$, stable fixed point {\bf P} can not be
reached from the initial coupling values $u>0$, $v<0$ (this region
as well as other unphysical for RIM regions are shown in grey
tone), pure Ising fixed point {\bf I} is unstable whereas random
fixed point {\bf R} is both stable and accessible (stable fixed
points are shown by boxes).  \label{figfp} }
\end{figure}

The effective Hamiltonians (\ref{6}) and (\ref{8}) possess different
global variables: although the space dimension is the same ($d=3$)
the symmetry and order parameter components number differ. Thus
one may expect that by application of renormalization group
approach they will lead to different critical behaviour. One of
the central notions in the renormalization group formalism applied
to the critical phenomena is the notion of the fixed point of the
renormalization group transformation. If a fixed point
exists and it can be reached from the initial values of the couplings
it corresponds to a critical point of the system. Applying the
renormalization group transformation to the effective Hamiltonian
(\ref{6}) and starting from positive values of $u$ one reaches the
stable fixed point $u^*$ which corresponds to the critical point of a
$d=3$ pure Ising model and leads in particular to the results
(\ref{4}). The fixed point structure for the effective Hamiltonian
(\ref{8}) is sketched in figure \ref{figfp}. The earliest qualitative
results about the structure of RIM fixed points appeared in
mid-seventies \cite{Khmelnitskii75,Lubensky75,Grinstein76}. Later
analysis supported this picture \cite{Sokolov77,Newman82,note3}:
indeed the fixed point {\bf I} of the $d=3$ pure Ising model appears
to be unstable and a new stable fixed point {\bf R} exists (see figure
\ref{figfp}). Thus the general answer of the renormalization group
analysis of the RIM supports non-perturbative results of Refs
\cite{Harris74,Chayes86}: RIM critical behaviour is governed by
critical exponents different from those of the pure Ising model.

In the subsequent chapters  we will have a look how this statement
was made clear in experimental, Monte Carlo and theoretical
studies.


\section{Experimental study}
\label{III} The Ising model is represented in experiments on
magnets by crystalline difluoride of a transition metal, normally
iron or manganese. In ${\rm FeF_2}$ which possesses a rutile
crystalline structure with $a=4.697 \AA$ and $c=3.309 \AA$ at room
temperature \cite{Hutchings72}, the spins of metal ions align
along $c$ axis in such a way that spins of the body-centered ions
are oriented in opposite directions to those of the corner ions.
The system can be well described  by  a Heisenberg--like spin
$S=2$ Hamiltonian with quadratic anisotropic single ion terms and
other interactions less than $6 \%$ of interspin exchange force
constant $J=0.45$ meV, as follows from the spin-wave dispersion
relations on the basis of neutron scattering measurements
\cite{Hutchings70}. Dominant intersublattice exchange interaction
and high anisotropy makes ${\rm FeF_2}$ a very good experimental
realization of an Ising antiferromagnet, where the order parameter
is the sublattice magnetization.   Experimental studies confirm
pure Ising model critical behaviour for the reduced temperature
range $|\tau|<10^{-1}$ ($\tau \equiv (T-T_N)/T_N$) around N\'eel
temperature $T_N$. Another example is ${\rm MnF_2}$ with similar
crystalline structure but much weaker single ion anisotropy than
${\rm FeF_2}$, however the experimental study showed the substance
to belong to the Ising universality class
\cite{Heller66,Dietrich69,Schullhof70} as well.

A material which corresponds to RIM can be obtained as a
crystalline mixture of two compounds on the basis of ``pure''
Ising model matrix (see Table \ref{table_experiment}).  A
corresponding site-diluted uniaxial alloy can be prepared by
substitution of ${\rm FeF_2}$ (${\rm MnF_2}$) with non-magnetic
isomorph ${\rm ZnF_2}$. Experiments on the critical behaviour of
random systems are extremely sensitive to the sample quality.
Asymptotic critical behaviour is observable only very close to
$T_N$ and macroscopic non-statistical gradients in concentration
cause variation of $T_N$ through the sample smearing out the sharp
transition. Therefore, to provide a satisfactory good realization
of random substitutional disorder of magnetic ions (${\rm
Fe^{+2}}$, ${\rm Mn^{+2}}$) by non-magnetic ones (${\rm
Zn^{+2}}$), mixed crystals ${\rm Fe_pZn_{1-p}F_2}$ as well as
${\rm Mn_pZn_{1-p}F_2}$ should be grown with very high crystalline
accuracy of the mosaic pattern, chemical homogeneity and
especially with small impurity concentration gradients. The last
can be achieved by choosing a concentration of impurities such
that \cite{Belanger82} $d T_N/d p = 0$  or  by a geometry which
takes into account that usually gradients are parallel to the
growth axis.  Early experimental studies on the disordered
crystals critical behaviour \cite{Meyer78,Cowley80} proved the
crucial role of a sample quality and high quality samples allowed
to observe sharp phase transition and to measure the dependence of
N\'eel temperature $T_N$ on $p$ via linear birefringence method
\cite{Belanger80}. This provided a basis for universal critical
properties measurements. The earliest experimental study of the
critical exponents governing the sharp transition at a weak
quenched dilution was the nuclear magnetic resonance measurement
of magnetization in ${\rm Mn_{0.864}F_{0.136}F_2}$
\cite{Dunlap81}. The value of magnetization exponent $\beta$ was
found to differ strongly from that in non-diluted sample (see
Table \ref{table_experiment} for details).

In a few years the study of Ref. \cite{Dunlap81} was corroborated
by  nuclear scattering (NS) measurements in a two-axis
spectrometer configuration \cite{Birgeneau83} of the staggered
susceptibility and the correlation length in a crystal on the iron
basis with concentration $p=0.5$ in the high dilution regime. In
the experiment the smearing effect was eliminated by masking the
crystal to expose only a small homogeneous region, while the
relative temperature control to about $0.01K$ together with the
substance N\'eel temperature $42.50 K$ gave the accuracy $\delta
|\tau| \sim 5\times 10^{-4}$ for reduced temperature. The data for
the inverse correlation length and the inverse staggered
susceptibility were fit to power laws with critical amplitudes and
exponents as free parameters. The power law was shown to be
satisfied well within $10^{-3} \div 2 \times 10^{-1}$ with the
same critical exponents (see Table \ref{table_experiment}) below
and above $T_c$ and differing strongly from those of the pure
Ising model. To support the result by alternative experimental
methods caloric properties were measured. Utilizing the fact that
for transition--metal difluorides the temperature derivative of
the linear birefringence is proportional to the magnetic specific
heat within \cite{Belanger80,Gehring77} $|\tau| < 10^{-1}$,
measurements of the critical exponent $\alpha$ were performed for
a sample ${\rm Fe_{0.6}Zn_{0.4}F_2}$ with $T_N=47.05$. In order to
minimize the effect of concentration gradients the laser beam was
oriented perpendicularly to the concentration gradient. The
numerical value of the critical exponent was extracted by fitting
the data to the temperature integral of specific heat scaling
function taking into account correction-to-scaling terms (i.e. the
first two terms in the expansion (\ref{3})). In contrast to the
pure Ising case, the data obtained  yielded $\alpha<0$ (see Table
\ref{table_experiment}). Within no region in reduced temperature
$\tau$ any evidence of pure Ising behaviour or a crossover from
pure to random fixed point critical behaviour was found. So one
concluded that the crossover is either outside the critical region
or it is too slow. The data obtained in Ref. \cite{Birgeneau83} on
the  basis of two different substances and quite different
experimental procedures  proofed that the scaling relation
$d\nu=2-\alpha$ holds for them and thus supported strongly scaling
in dilute systems.

\begin{figure}[htbp]
\epsfxsize 100mm
\centerline{\epsffile{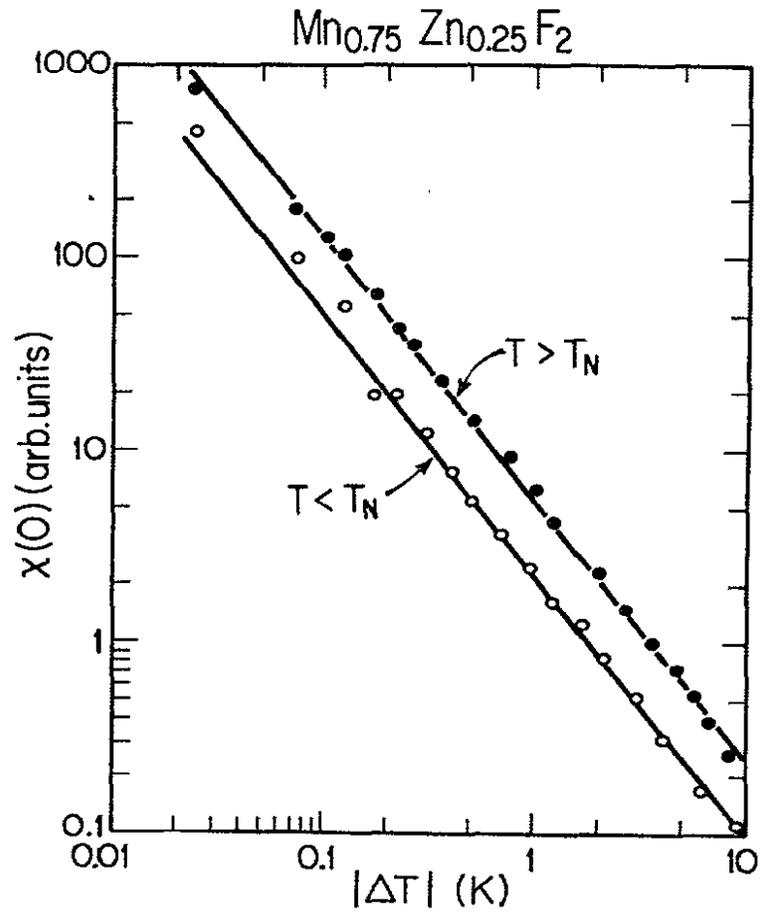}}
\vspace*{5mm}
\caption{\label{figmitchell1}
Neutron scattering measurements of the susceptibility $\chi(0)$ in
${\rm Mn_{0.75}Zn_{0.25}F_2}$ \protect\cite{Mitchell86}. The solid lines are
the results of fits to single power
laws with exponents $\gamma\simeq1.364$ above and below N\'eel temperature
$T_N$. The critical behaviour is governed by RIM asymptotic critical
exponents over the reduced temperature interval
$4 \cdot 10^{-4}<|\tau|<2\cdot10^{-1}$.
The figure is taken from the Ref. \protect\cite{Mitchell86}.
}
\end{figure}

\begin{figure}[htbp]
\epsfxsize 100mm
\centerline{\epsffile{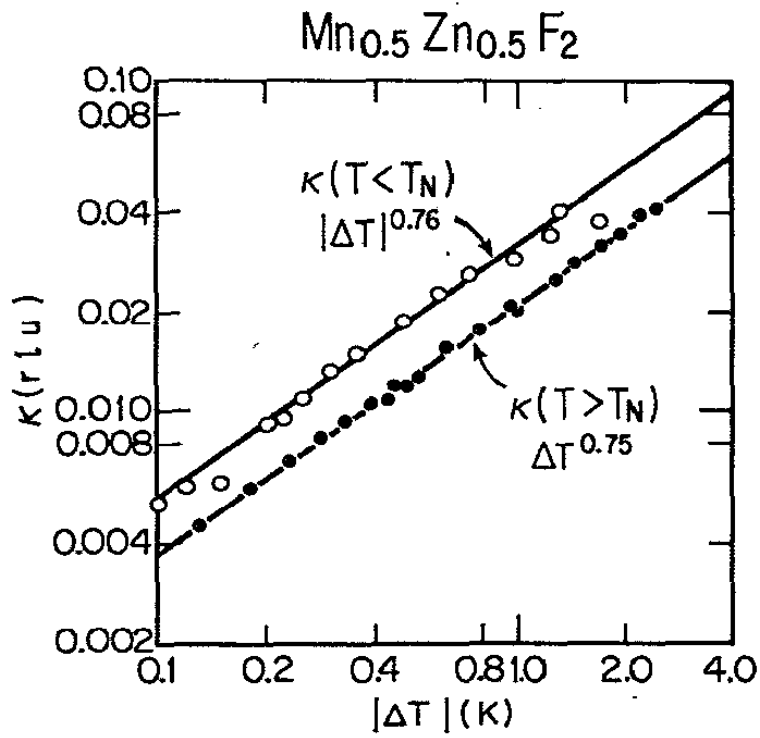}}
\vspace*{5mm}
\caption{\label{figmitchell2}
Neutron scattering measurements of the longitudinal inverse correlation length
$\kappa$ in ${\rm Mn_{0.5}Zn_{0.5}F_2}$ above and below $T_N$
\protect\cite{Mitchell86}. The solid lines are
the results of fits to
laws with the exponents $\nu'\simeq0.76$ for $T<T_N$ and $\nu\simeq0.75$
for $T>T_N$.
The data give $\nu'=\nu$ within the errors (see Table
\protect\ref{table_experiment}). The law holds in the reduced temperature
interval $5 \cdot 10^{-3}<|\tau|<10^{-1}$.
The asymptotic region still is not reached as the observed effective
exponents values show. The figure is taken from the
Ref. \protect\cite{Mitchell86}.
}
\end{figure}

With a typical energy of the neutron beam about $10$ meV, the NS
measurements appeared to be one of the most useful method to study
criticality, however they turned out to be in particular very
sensitive to the sample quality since typical sample sizes are
less than several millimeters. The NS measurements contain
contributions from transverse and longitudinal spin fluctuations,
which have to be separated in a detailed data analysis. This
explains why in the neutron scattering measurements of the Ref.
\cite{Belanger86} which were performed on ${\rm Fe_pZn_{1-p}F_2}$
with concentration $0.46$ on the similar device, special attention
was paid to the quality of a sample.  In the study a variation $2
\times 10^{-4}$ in concentration over the entire volume was
achieved which permitted scattering studies up to $|\tau| \geq
10^{-3}$. Within the interval $1.5 \times 10^{-3} \leq |\tau| \leq
10^{-1}$ the inverse correlation length was obtained from the
width of the Lorentzian fits as a function  of temperature, and
the correlation length exponent was extracted from the power-law fits.
The fit to the power ansats for $\chi$ with background term yielded
$\gamma$ after extrapolation to the wave vector length $q=0$ (see
Table \ref{table_experiment}). An alternative substance with the
base Mn (${\rm Mn_pZn_{1-p}F_2 }$) was studied by neutron
scattering experiments a year later \cite{Mitchell86}. The sample
with $p=0.75$ had overall spread in concentration $0.001$ and
allowed to perform measurements up to $|\tau| \sim 4 \times
10^{-4}$ (see figure \ref{figmitchell1}), while the $p=0.5$ sample
was of lower quality with a spread in concentration of $0.005$.
Quite high quality of the samples as well as the temperature
control up to $0.05$ K allowed to obtain critical amplitudes
ratios and exponents with good accuracy. However, systematic
errors, which were assumed to be due to the resolution
corrections, the quasi-elastic approximation, concentration
fluctuations and background effects, appeared to be of more
importance than statistical ones. The authors stressed that since
no correction-to-scaling terms were used, the exponents found are
effective ones. However the $p=0.75$ sample was shown to exhibit
RIM critical behaviour over the reduced temperature interval $4
\times 10^{-4} < |\tau| < 2\times 10^{-1}$, whereas the critical
exponents at higher dilution $p=0.5$ had not reached their
asymptotic RIM values (see Table \ref{table_experiment} and figure
\ref{figmitchell2}).

The results  mentioned above for the single crystal of transition
metal fluorides were corroborated  \cite{Hastings85} by neutron
scattering technique for a sample of dysprosium aluminum garnet
${\rm Dy_3 Al_5 O_{12}}$ powder. A powder was prepared of this
cubic non-collinear Ising antiferromagnet together with
non-magnetic Yttrium to avoid gradients in composition which
unavoidably occur. The critical exponent $\beta$ increased in
disordered sample in comparison to the value of the ``pure'' Ising
model (see Table \ref{table_experiment}) and provided an extra
evidence of the mapping between experiment and theoretical
insight.

It is known that  static critical behaviour of the diluted Ising
magnet in a uniform field $H$ along the uniaxial direction behaves
according to the random-field Ising model
\cite{Fishman79,Cardy84}. The random-field Ising model is the
subject of intensive recent experimental studies (see
\cite{Belanger91,Belanger00} for a review). Two studies on the NS
measurements in RIM appeared in the middle of 90th in the context
of random-field model investigations \cite{Belanger95,Belanger96}.
In the NS studies the problem of the Bragg scattering saturation
due to extinction effects was reduced by growing a $3.4$ $ \mu m$
epitaxial film of ${\rm Fe_{0.5}Zn_{0.5}F_2}$ on a $(001)$ ${\rm
ZnF_2}$ substrate. The small $X$-ray rocking curve line-width of
the $(002)$ reflection showed the film to be of very high quality
which was in part due to the nearly identical $a$-axis lattice
parameters of the substrate and substance. Nevertheless, the film
contained $10^4$ lattice spacings and could be considered as a
three-dimensional object as was proven in  ${\rm FeF_2}$ showing
the critical behaviour of the ``pure'' $d=3$ Ising magnet
\cite{Belanger93}. The transverse $(100)$ Bragg scattering scans
data appeared to be well described by a Gaussian distribution with
a background term. Fitting the Bragg amplitude vs temperature to
the simple power law a critical exponent $\beta=0.35$ was found
for $H=0$, however, a rounding of the expected critical behaviour
of the magnetization close to $T_N$ prevented an accurate analysis
of the data. To reveal whether this was caused by poor sample
quality the measurements on the $3.4$ $ \mu m$ film of ${\rm
Fe_{0.52}Zn_{0.48}F_2}$ were performed \cite{Belanger96}. The
higher resolution data confirmed the value of $\beta$ obtained and
showed that rounding was only due to resolution. Recently in a
thin crystal of $0.44$ mm the  scattering measurements were
extended  from  $|\tau| \geq 10^{-2}$ to $|\tau| \geq 10^{-4}$ to
obtain the correlation length and susceptibility critical
exponents \cite{Slanic98,Slanic98a,Slanic99}.

The M\"osbauer spectroscopy (MS) studies of the critical behaviour
started in 1986 on a class of ${\rm Fe}$-based substances with
various magnetic atom concentrations \cite{Barret86}. The method
was tested earlier for the ``pure'' Ising antiferromagnet
 ${\rm FeF_2}$, where the sublattice magnetization was shown to be
proportional to the iron hyperfine field $h$ which can be measured
in MS \cite{Wertheim67}. To apply the approach to a dilute magnet,
the variation of the concentration of ${\rm Zn}$ was reduced to
$10^{-4}$ and a temperature stability of $0.002$ K was achieved.
In order to fix the critical temperature the values of $\beta$
were chosen such that the plot $h(T)^{1/\beta}$ vs $T$ became a
straight line intercepting the abscissa axis at $T_N$. This method
appeared to be not sensitive to particular values of $\beta$ and
gave the accuracy $0.05$ K for critical temperature location.
Though the critical exponent $\beta$ can be obtained from the
curve slope in double-logarithmic plot, this method produced
results only after separating the data into two intervals in
$\tau$ which were characterized by obviously different critical
exponents: one of the ``pure'' Ising model and the other one of
RIM. The crossover in  $\beta$ occured within a very narrow range
and at relatively  large values of $10^{-1} \geq |\tau| \geq
10^{-3}$ for small dilution $p \geq 0.93$. The subsequent MS study
appeared two years later \cite{Rosov88} and is characterized by a
very detailed analysis of the data. An advantage was the high
quality of the sample grown from a stoichiometrical mixture of
${\rm FeF_2}$ and ${\rm ZnF_2}$ powders, each prepared by reacting
a metal sponge with dry ${\rm HF}$ at $+900^{o}$. In the
experiment concentration gradients in a single crystal absorber
were minimized by choosing the direction of the $\gamma$-ray
parallel the growth axis, both perpendicular to the plane of the
sample disk of exposing sizes $4 \times 5 \times 0.1$ mm. 20
constant-acceleration spectra were obtained within $3 \times
10^{-4} < |\tau| < 0.86$. Very good temperature stability $3$
mK/day and account of correction-to-scaling permitted to obtain
the asymptotic RIM magnetization critical exponent $\beta$ from
the data in the investigated interval (see Table
\ref{table_experiment}).

The optical linear birefringence  method (LB) introduced in Ref.
\cite{Birgeneau83} for the study of critical  behaviour of the
magnetic part of the specific heat $c_m$  in disordered magnets was
continued in Refs
\cite{Ramos88,Ferreira91,Slanic98}. The proportionality between
the temperature derivative ${\rm d} (\Delta n)/{\rm d} T$ of the
optical birefringence $\Delta n$   and $c_m$ is assumed for
studies of optically transparent materials. While the caloric
properties in the critical region can be measured also by the pulsed
heat technique, the LB method has several advantages: first of all,
the non-magnetic contribution to the temperature derivative of the
optical birefringence  is insignificant,  contrary to thermal
techniques where the non-magnetic phonon background is often large
and difficult to eliminate; secondly, one can minimize the
concentration fluctuations effects by applying a laser beam
perpendicularly to the concentration gradients.
In the first experiment \cite{Ramos88}
a class of single crystals with various impurity concentration
showed the exponent $\alpha$ to be independent
of concentration and in very good agreement with the theoretical
predictions (see Table III), while the two subsequent
studies \cite{Slanic98,Ferreira91} showed the equivalence of the LB method
results with the direct heat pulsing measurements. Magnetic $X$-ray
scattering successfully applied to the ``pure'' Ising crystal
\cite{Goldman87} was applied to the ''random'' Ising crystal for high
dilution in Ref. \cite{Thurston88}.  For samples of sizes $5.7 \times 6.4
\times 8.9$ mm and of good crystallographic as well as mosaic quality  the
mean-field value for the exponent $\beta$ has been found up to $|\tau| \leq
0.06$, while closer to $T_N$ an  exponent $\beta$ consistent with
the RIM value was obtained.


\section{Monte Carlo simulations} \label{IV}

The power--law singularities like (\ref{2}) in physical
quantities only appear in the thermodynamic limit when the system volume
and the number of particles tend to infinity. Therefore an obvious obstacles
in  computer ''experiments'' is that simulations can be done for a
system of finite size  only. Moreover the asymptotic temperature interval
fails to be reached since $T_c$ itself is not known exactly,
whereas the location of the critical temperature is crucial
for the accuracy of critical exponents determination.
One can extrapolate to the thermodynamic limit assuming that the
regime of a constant asymptotic critical exponent is established
already starting from a certain finite system size and within an
interval around $T_c$ of a certain non--zero size. The last
assumption exploited in computer simulations leads to a very
narrow temperature intervals below and above $T_c$ reliable for
data sampling. The upper bounds are the temperatures where
correction-to-scaling accounts for, while the lower bounds follow
from finite-size effects and are of order $10^{-3}$ for typical
system sizes achieved now in the Monte Carlo (MC) simulations.
Therefore, the exponents obtained in the manner described
often are effective critical exponents which
characterize the critical behaviour in the observable temperature
range. On the other hand different time scales govern the
dynamical behaviour of pure and disordered spin systems and only
very long-running simulations yield reliable data.
The relaxation time
increases drastically not only when approaching $T_c$. Moreover, in
the early simulations it increased also with dilution and
yielded in several decades greater values when passing from a pure
$p=1$ system to a diluted one with $p=0.6$.
However an application of a more elaborated simulation technique
resulted in an opposite behaviour: the decrease of the relaxation time
for a certain disordered sample with an increase of $p$ \cite{Berche}.
And it is the configurational averaging that leads to overall increase
of the computation time. The
statistical errors for thermodynamic observables result mainly from
variance in configurational space and are larger than usual
statistical errors in finite-size simulations with usual now
statistics of $10^6$ MCS (Monte Carlo steps per spin).
Consequently, the accuracy is bounded by the available CPU time
from one side and limited by the number of samples to perform
the configuration average of a disordered model \cite{Brout59}
from the other side. Efficient algorithms together with high-speed
computers partially solve the task.

Despite of this unpleasant situation Monte Carlo studies
provided deep insight into the origin of the phase transition in
the RIM as well as resulted in reliable numerical values
for the critical exponents. MC studies of three dimensional RIM
systems have been carried out over the last two decades
\cite{Landau80,Marro86,Chowdhury86,Braun88,Braun89,Wang89,Wang90,%
Holey90,Heuer90,Heuer93,Prudnikov93,Hennecke93,Parisi98,Wiseman98a,%
Wiseman98b,Aharony98,Marques00,Marques00a,Marques00b,Hukushima00}.
The first search \cite{Landau80} for universal critical
characteristics of a RIM was performed on a simple cubic lattice
of a size up to $30^3$ on the basis of an importance sampling MC
method \cite{Binder79}. The positions of specific heat peaks
corresponding to finite-lattice pseudo-critical temperature were
extrapolated by means of finite-size scaling theory to $T_c$, and
the accuracy of $500-5000$ MCS per data point with averaging over
several different starting configurations of impurities did not
allow to discriminate between disorder relevance or irrelevance.
From a fit the value of the order parameter  exponent $\beta$ and
the susceptibility exponent $\gamma$ of RIM could not be
distinguished within the error bars from the corresponding
exponents of the pure (non-diluted) system for any dilution (see
Table II, Ref. \cite{Landau80}).  As a possible explanation it was
stated that the lattice studied were too small to reach the
critical region. The predictions of the authors that ``impure''
critical behaviour can be observed only for systems with larger
critical regions (larger $\alpha_{pure}$) was objected by the MC
simulations of RIM  on larger lattices with narrower dilution
range \cite{Marro86}, higher statistics of $5000\div11000$ MCS,
larger number of averaged samples and more detailed analysis of
simulational data. The phase transition temperature $T_c$ was
located by adjusting data of specific--heat and numerical
derivative of the energy. In this way the critical temperature
accuracy was improved to $0.004$ and the relevance of disorder for
the universal critical properties was concluded.  This result was
confirmed by magnetization data, though the critical exponent
$\beta$ was found to vary continuously with the magnetic sites
concentration and no temperature region with constant $\beta$ was
found. Similarly to Ref. \cite{Landau80} the authors of Ref.
\cite{Marro86} stressed that while they cannot exclude the
existence of a tiny impure critical region it must be unobservable
for the weak dilution experimentally as well as by Monte Carlo
methods, and effective critical exponents varying continuously
with dilution would always show up according to the two fixed
points scenario (see fig. \ref{figfp}). Even the possibility of
the existence of a line of fixed points, one for each impurity
concentration, was not excluded (compare the dependence of the
magnetization critical exponent $\beta$ on the concentration of
magnetic sites in Table \ref{table_MC}, Ref. \cite{Marro86}).

The results mentioned above appeared to be in a good agreement
with the MC
data of Ref. \cite{Chowdhury86} obtained on the basis of a
multi-spin coding program where twice as large systems and $8000$
MCS statistics were used for a given concentration and
temperature. The effective exponent $\beta$ was found to increase
continuously with dilution as well. However this was believed to be a
consequence of the fact that one needs to compute the equilibrium
magnetization much closer to the critical point. Two years later
the conclusion that reliable critical exponent values could not be
obtained by the simulation of systems with sizes $l \leq 20$ was
drawn in Ref. \cite{Braun88}. Statistical errors in the
determination of the critical temperature from susceptibility and/or
specific heat data, together with the extrapolation by finite-size
scaling were thought to be responsible for this situation.
In particular correction-to-scaling \cite{Barber85}
should be taken into account. Instead, the simulation with the
statistics $5000\div11000$ MCS, but no configuration average
allowed a determination of the exponent $\beta$ obviously larger
than the pure one, while the determination of $\gamma$ were not
done because of a large scatter of the data.
A reanalysis of the previous MC simulations
\cite{Landau80,Marro86,Chowdhury86} led
the authors to the conclusion that a large
part of critical regime is dominated by the crossover from the pure
Ising exponents to RIM ones: the results of Ref.
\cite{Landau80} may correspond to a plateau value of the effective
critical exponent governed by the pure Ising fixed point, while
refined results of Refs \cite{Marro86,Chowdhury86} correspond to
an intermediate regime before the asymptotic regime, unaccessible
near $T_c$, is reached.  Enforcing the concept of effective critical
exponents \cite{Kouvel64,Riedel74} (see formula (\ref{2})) the
authors of Ref. \cite{Braun88} claimed that  the effective
exponents they found were in  qualitative agreement with the
renormalization group flows (see fig. \ref{figfp}).

Soon it became clear that the  concentration dependence
of the critical exponents observed previously in MC simulations
may be also due to large relaxation time at
criticality. The Swedsen-Wang  algorithm applied to RIM  simulation of
Ref. \cite{Wang89}  resulted in the conclusion that a finite-size
analysis of susceptibility data
is inadequate to check the new critical exponents appearance for a RIM
since the ratio $\gamma/\nu$ does not change with the system size.
On the other hand, the determination of $T_c$ from
susceptibility maxima and the fourth-cumulant intersection method allowed to
calculate the effective critical exponent $\gamma$ by fitting the
data points in successive time intervals. In this way the
susceptibility and the correlation length critical exponents were found
to be independent of concentration  in a wide range of
dilution (see Table II, Ref. \cite{Braun88}).
As observed in Ref. \cite{Wang89} the exponent $\gamma$
appeared to be higher  than the theoretical asymptotic value.
Thus it has been argued that $\gamma_{eff}$ is a non--monotonic function of
reduced temperature with a maximum value, the larger the bigger  the
dilution. For the   particular case of magnetic sites concentration
$0.8$ the prediction was
investigated in Ref. \cite{Wang90} by the Swedsen-Wang algorithm and the
single-cluster generalization of Wolf for a rather large system. While the
$\gamma/\nu$ ratio confirmed the result of Ref. \cite{Wang89}, the
effective critical exponent $\gamma$ did not show a maximum but
increased continuously approaching $T_c$.

The assumption that the interaction among block-spins in diluted
systems may be represented by renormalized couplings of pure
systems was used in the MC renormalization group approach for RIM
in the weak dilution region ($p \geq 0.8$) \cite{Holey90}. The
values for the correlation length critical exponent $\nu$ were
found to depend on the concentrations of the magnetic sites. For
$p=0.8$  $\nu$ satisfied the exact inequality $\nu > 2/d$  for a
diluted system \cite{Chayes86} and this led the authors of Ref.
\cite{Holey90} to the assumption that they obtained  the
asymptotic value of the correlation length critical exponent. They
subsequently estimated the width of the asymptotic critical region
for this dilution ($|\tau|_{\rm crit} \leq 3 \cdot 10^{-4}$) where
no influence of the pure Ising model fixed point is seen.
Alternatively, for $p=0.9$ the inequality \cite{Chayes86} is not
satisfied for $\nu$ and either the critical region $|\tau|_{\rm
crit} \leq 1.3 \cdot 10^{-4}$ was not reached and thus the
critical behaviour is that of pure Ising model, or the crossover
to the RIM asymptotical critical behaviour for $p=0.9$ is
observable only in larger systems.  For $p=0.4$ the accuracy of
data obtained was too poor to find any values.

\begin{figure}[htbp]
\epsfxsize 100mm
\centerline{\epsffile{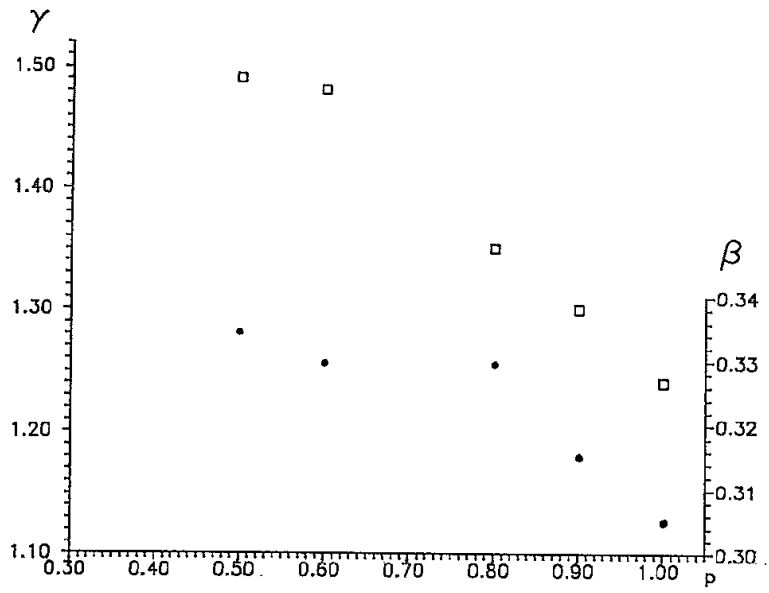}}
\vspace*{5mm}
\caption{\label{figheuer} Obtained in MC simulations
\protect\cite{Heuer90} RIM effective critical exponents $\beta$
(dots) and $\gamma$ (squares) for different dilutions $p$. All
exponents change continuously with dilution and are clearly
different from their pure system values. The figure is taken from
Ref. \protect\cite{Heuer90}. }
\end{figure}

Precise MC data of Ref. \cite{Heuer90} became available by
application of the new cluster algorithm \cite{Swedsen87} and a
refined vectorized  implementation of local algorithms \cite{Heuer90a}.
A progress in MC studies occurred when an improved version of the
multi-spin coding program appeared \cite{Heuer90a,Heuer90}. It allowed to
verify all previously obtained simulation data for the RIM  with
the unprecedented statistics of up to $3\times10^5$ MCS and 10 averaged
configurations for systems with $p \geq 0.8$ and up to
$1.2\times10^6$ MCS for higher dilution. For instance, in the test
simulation the statistics allowed to obtain the susceptibility and
magnetization critical exponents of a $2d$ pure Ising model with an
accuracy of $1 \%$  and  $3\%$ respectively. For the RIM,  the
effective exponents $\gamma,\beta$ and $\zeta=1-\beta$ (the last
one describes the divergence of the magnetization-energy
correlation function) were obtained \cite{Heuer90} and were shown to be
concentration dependent in the concentration region $0.5 \leq p <
1$ (see Table \ref{table_MC}). The $T_c$ value as well as
$\gamma$ were obtained by fitting susceptibility data to a simple
power law; the $T_c$ location was then verified  by the 4th and 6th
cumulant of magnetization. The investigation of the
energy--magnetization correlation function allowed to check if
scaling holds; the critical exponents $\zeta$ and $\beta$ were obtained
by fitting the data to the simple power law with the value $T_c$ taken
from the susceptibility analysis.  All data showed power--law behaviour
within the chosen temperature range but the constant values of critical
exponents changed with concentration (see figure \ref{figheuer}).
For instance, $\gamma$ increased from the
pure Ising value and smoothly changed with dilution achieving plateau
value at $p=0.5$, $\beta$ similarly increases while $\zeta$
decreases in such a way that their sum equals $1$ within errors.
The conclusion of Ref. \cite{Heuer90} states that while new
critical exponents changing with the dilution were observed, no
line of stable fixed points exists as supposed e.g. in \cite{Marro86}.
The result originates from a crossover from pure to diluted and
percolation to diluted regimes with effective exponents for all
concentrations.

The evidence of crossover phenomena  motivated the
authors to undertake a more systematic study. The data of Ref.
\cite{Heuer90} were revised three years later in Ref.
\cite{Heuer93}  where a cumulant method  was used in order to
determine $T_c$. For thermal averaging up to $40000$ MCS were
performed and for the configurational averaging up to $32$
configurations were used (compare this with $5000$ MCS and only
several  samples for the configurational average in Ref.
\cite{Landau80}). The critical exponents $\zeta$,  $\beta$ and
$\gamma$  were shown to vary with concentration in consistency
with the two fixed point scenario for the weak dilution. However
for the strong dilution  ($p=0.5,0.6$)  an influence of another
``percolative'' fixed point was assumed. A central part of Ref.
\cite{Heuer93} dealt with the magnetization-energy scaling
function. It was shown that dilute systems have a complex
crossover behaviour before they reach their asymptotic  critical
region with values for the exponents consistent with the weak
random fixed point. For instance, pure systems reach their
asymptotic limit for small system size. In weakly random
systems($p\geq0.8$) the asymptotic values for the exponents are
smoothly approached  from below whereas in strongly diluted
systems from above. This happens at the characteristic system size
$\lambda_p$ depending on the concentration $p$.  $\lambda_p$ was
estimated as $20\div30$ lattice constants for pure, $50$ for
$p=0.95,0.9$ and about $100\div150$ for $p=0.6$; at the
percolation point $\lambda_p$ diverges. Thus more strongly
disordered systems necessitate a refined analysis with an
appropriate treatment of their percolative structural effects
relevant for non-asymptotic sizes and correlation lengths.
Accidentally the crossover function changes its sign at $p \sim
0.8$ so the asymptotic values appear to be reached already for
small systems (c. f. Refs \cite{Wang90,Heuer90}).

Due to works \cite{Wang89,Wang90} and especially
\cite{Heuer90,Heuer93} it became clear that the concentration
dependent critical exponents found in MC simulations are effective
ones, characterizing the approach to the asymptotic region. This
point of view found its support in a conjecture about a step-like
universality of the 3d diluted magnets \cite{Prudnikov93}. An
attempt to study the diluted model at $p=0.6$ by means of taking
MC data from different subsystems of one large system was made in
Ref.  \cite{Hennecke93}.  It was concluded  that  the
resulting values of the critical  exponents  are strongly
influenced by the size of the system.

\begin{figure}[htbp]
\epsfxsize 100mm
\epsfysize 70mm
\centerline{\epsffile{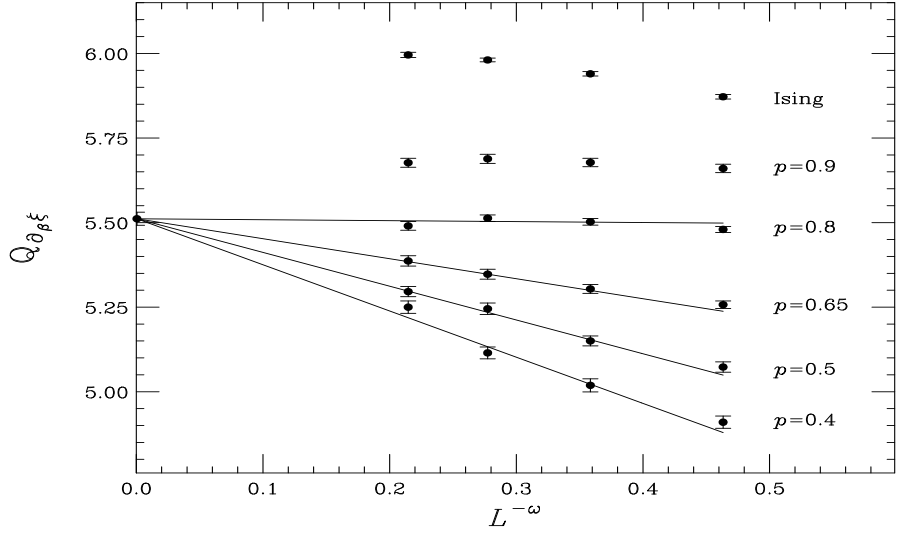}}
\vspace*{5mm}
\caption{\label{figparisi} Determination of the
correction-to-scaling exponent $\omega$ in the MC simulations  of
Ref. \protect\cite{Parisi98}. Quantity
$Q_{\partial_{\beta}\xi}=2^{1+1/\nu}$ is plotted for different
dilutions $0.4\leq p\leq 1$ and lattice sizes $8\leq L\leq 128$.
One can see that at $p=0.9$ the system is crossing over from the
pure Ising fixed point to the diluted one even for $L=128$. The
solid lines correspond to a fit $\omega=0.37$ yielding the same
infinite volume extrapolation for all $p\leq 0.8$. The figure is
taken from Ref. \protect\cite{Parisi98}. }
\end{figure}

Recently, the critical behaviour of the RIM was reexamined by MC
method in Refs \cite{Wiseman98a,Wiseman98b} for
concentrations $p=0.8, 0.6$  and in Ref. \cite{Parisi98} for $0.4
\leq p \leq 0.9$ respectively. In particular, the simulations
\cite{Wiseman98b} revealed that a disorder realized in a canonical
manner (fixing the fraction of magnetic sites) leads to different
results in comparison with disorder realized in a grand-canonical
manner (see Table \ref{table_MC}). Studies of Ref.
\cite{Parisi98} were based on the importance of taking into
account the leading correction-to-scaling term in the infinite
volume extrapolation of the MC data and thus the analysis does not
agree with the data of Ref. \cite{Wiseman98b}. The results of
the simulations for the concentrations $p=0.9,0.8,0.6,0.4$ were
extrapolated to infinite system size (see figure \ref{figparisi})
and lead to the proof  of universal critical behaviour of the
site-diluted Ising model in a wide range of concentrations. In
particular the value of the correction-to-scaling exponent
$\omega$ was found to be $\omega=0.37\pm 0.06$ which is almost
half as large as the corresponding value in the pure $d=3$ Ising
model  \cite{note1}
 $\omega=0.799\pm 0.011$. The smaller the  value of
$\omega$  the larger the interval where it has to be taken into
account (c.f. formula (\ref{3}) of the chapter \ref{I}). Thus the
smallness of $\omega$ in the dilute case explains its importance
for an analysis of the asymptotic critical behaviour.

Another important question considered in Refs
\cite{Wiseman98a,Wiseman98b} is the problem of self-averaging in
RIM. The Gibbs approach to the static collective phenomena rests
on the statistical independence of macrosamples according to the
short range nature of inter-particle interactions. In agreement
with this approach any thermodynamic extensive quantity $M$ is
(strongly) self--averaged. It means that the normalized square
width $R_M$ of the squared variance of its sub-sample values
behaves as $R_M=V_M/M^2 \sim 1/n \sim l^{-d}$, where $n$ is the
number of sub-samples and $l$ is the system linear size. However,
in the vicinity of the critical point the statistical independence
does not hold since the correlation length of a system $\xi$ can
be arbitrary large $\xi \sim l$, and thus subsystems cannot be
considered as independent. The concept of weak self-averaging
takes account of the case when a number $x_1, 0<x_1<d$ exists such
that $R_M$ at criticality scales as $l^{-x_1}$.
In contrary, if $R_M \rightarrow {\rm const} \not = 0$,
$M$ is called to be not self--averaging. It has been predicted
on the basis of heuristic arguments that for random models all
extensive quantities are strong self-averaging far from
criticality. Yet for a quantity $M$ scaling at the critical point
as $l^{\rho}$ the strong self--averaging should fail. Here the
squared variance $V_M$ is expected to scale as
$l^{2\rho+\alpha/\nu}$ for $\alpha_{random}<0$ implying $R_M \sim
l^{\alpha/\nu}$ or weak self-averaging \cite{Wiseman95}.

Lack of self--averaging in RIM is not only of high theoretical
interest. The reliability of $MC$ simulations depends on the
answer whether the increase of the lattice size improves the
statistics of the simulations. If a quantity is not
self--averaging the simulational data are unreliable. Theoretical
studies based on the renormalization group approach confirmed the
strong self--averaging for $l \gg \xi$. In contrast, special MC
investigations in finite systems found no self--averaging in the
case of relevant disorder $\alpha_{_{pure}}>0$, while weak
self-averaging appeared to be the case only for irrelevant
disorder \cite{Aharony96}, in disagreement with Ref.
\cite{Wiseman95}. The  MC simulations of the Refs
\cite{Wiseman98a,Wiseman98b} were performed in order to {resolve
this puzzle.} It was shown that the normalized square width $R_M$
goes to a constant for large $l$, which is {\it independent} from
the dilution of grand--canonical type, while this is not the case
for the canonical type \cite{Wiseman98a}. The last result,
however, may have its explanation in the very slow approach of $R$ to
its universal asymptotic value in the case of canonical
realization of $3d$ random Ising model, estimated as
$l^{\alpha/\nu}$ \cite{Aharony98}.

The evolution of the self-averaging from pure Ising model to RIM
has been studied recently \cite{Marques00} in order to determine
the transition zone between pure and diluted Ising models'
universality classes. It was shown that the transition zone is
smoothly dependent on concentration of magnetic sites and
independent of lattice size. In contrary, critical exponents
values did not depend on concentration and were found identical to
previous data of Ref. \cite{Parisi98}. The universal value of the
normalized square width of the susceptibility in the
infinite-volume limit was estimated to equal
$R_{\chi}(\infty)=0.155$.

Apart from short-range site-dilution other realizations of
disorder have become a subject for MC simulations recently.
Thermally diluted Ising  model has been studied as a
generalization of the RIM \cite{Marques00a,Marques00b}.
There, the realization of  the
vacancy distribution is determined from the local distribution of
spins in a pure Ising model at criticality.  The critical
properties, in particular, the universality class of the model
appeared to differ strongly from the RIM one, however, they agreed
well with the theoretical predictions for long-range-correlated
disorder \cite{Weinrib83,note2}.

The problem whether the RIM fixed point also describes the phase
transition in the Ising model with random bonds has been studied
explicitly in Refs \cite{Hukushima00,Berche00}. Using a numerical
renormalization-group analysis the RG flows for random Ising
models have been obtained and the  existence of a fixed point
characterizing the random Ising model irrespectively to the type
of disorder has been shown  \cite{Hukushima00}.


\section{Renormalization group theory expansions}

\label{V} In the subsequent sections  \ref{V} and \ref{VI} we will
review results on RIM critical exponents obtained my means of
renormalization group (RG) methods. In section \ref{V} we report
the main relations of the field theoretical RG approach and dwell
upon perturbation expansion series available. Section \ref{VI}
deals with the RG series resummation methods and the results
obtained on their basis.
\subsection{Renormalization}
\label{V1} To describe theoretically the long-distance properties
arising in different systems in the vicinity of the second order
phase transition point it is standard now to use a field
theoretical RG approach \cite{rgbooks1,rgbooks2}. The
renormalization is used to remove divergences that occur during
evaluation of the bare vertex functions in the asymptotic limit.
For the RIM one-particle irreducible bare vertex functions are
defined as:
\begin{equation} \label{11}
\delta(q_1+\cdots+q_N)
\Gamma^N_0(q_1,\ldots,q_N;m_0,u_0,v_0;\Lambda_0)
= \int e^{i(q_1 r_1+\cdots+q_N r_N)}
\langle
\varphi(r_1)\cdots\varphi(r_N)
\rangle_{\rm 1PI}^{\cal H_{\rm RIM}}{\rm d}r_1 \cdots {\rm d}r_N,
\end{equation}
where the angular brackets denote statistical average over the
Gibbs distribution with the Hamiltonian (\ref{8}) in the replica
limit $n\rightarrow 0$ and the subscript $|_{\rm 1PI}$ indicates
that only one--particle irreducible diagrams are taken into
account. The functions depend on sets of momenta $q_1, \ldots,
q_N$ (with $\Lambda_0$ as a momentum cutoff) and the bare
parameters $m_0, u_0, v_0$ of the Hamiltonian (\ref{8}).
Divergences in (\ref{11}) occur in the asymptotic limit
$\Lambda_0\rightarrow \infty$. Their removing is achieved by a
controlled rearrangement of the series for the vertex functions.
Several asymptotically equivalent procedures serve this purpose.
For the purpose of this presentation we will use two complementary
approaches: {\bf (i)} dimensional regularization and the minimal
subtraction scheme \cite{tHooft72} and the fixed dimension
renormalization at zero external momenta and non-zero mass (a
massive RG scheme) \cite{Parisi80}.

Let us formulate relations of the renormalized theory. The renormalized fields,
mass and couplings $\phi,m,u,v$ are introduced by:
\begin{eqnarray} \label{11a}
\varphi &=& Z_{\phi}^{1/2} \phi, \\
m_0^2 &=& Z_{m^2} m^2 , \label{12} \\
u_0 &=& \mu^{\varepsilon} \frac{Z_{4,u}}{Z_{\phi}^2}u \, , \label{13} \\
v_0 &=& \mu^{\varepsilon} \frac{Z_{4,v}}{Z_{\phi}^2}v \, . \label{14}
\end{eqnarray}
Here, $\varepsilon=4-d$, $\mu$ is a scale parameter equal to the
renormalized mass at which the massive scheme \cite{Parisi80} is
evaluated  or in the minimal subtraction scheme \cite{tHooft72} it
sets the scale of the external momenta. $Z_{\phi}$, $Z_{m^2}$,
$Z_{4,u}$, $Z_{4,v}$ are the renormalizing factors. The
renormalized vertex functions $\Gamma_R^{N}$ expressed in terms of
the bare vertex function by:
\begin{equation} \label{14a}
\Gamma_R^{N}(q_1,\ldots,q_N;m,u,v) =
Z_{\phi}^{N/2} \Gamma_0^{N}(q_1,\ldots,q_N;m_0,u_0,v_0).
\end{equation}
are finite. This is the main content of the multiplicative
renormalizability  of the field theory defined by the Hamiltonian
(\ref{8}).

First let us consider {\em the minimal subtraction scheme}. Here,
the renormalizing $Z$-factors (\ref{11a})--(\ref{14}) are determined
by the condition that all poles at $\varepsilon=0$ are removed from
the renormalized vertex functions.
The RG equations are written bearing in mind that the bare
vertex functions $\Gamma_0^{N}$ (\ref{11})
do not depend on the scale $\mu$, and therefore their
derivatives with respect
to $\mu$ at fixed bare parameters are equal to zero. So one gets
\begin{equation} \label{15}
\mu \frac{\partial}{\partial \mu} \Gamma_0^{N}|_0=
\mu \frac{\partial}{\partial \mu} Z_{\phi}^{-N/2}
\Gamma_R^{N}|_0=0 \, ,
\end{equation}
where the index $|_0$ means a differentiation at fixed bare parameters.
Then the RG equation for the renormalized vertex function
$\Gamma_R^{N}$ reads:
\begin{equation} \label{16}
\Big ( \mu \frac{\partial}{\partial \mu} +
\beta_u \frac{\partial}{\partial u} +
\beta_v \frac{\partial}{\partial v} +
\gamma_{m} m  \frac{\partial}{\partial m} -
\frac{N}{2} \gamma_{\phi} \Big )
\Gamma_R^{N}(m,u,v,\mu) =0,
\end{equation}
and the RG functions are given by
\begin{eqnarray} \label{17}
\beta_u(u,v) &=& \mu \frac{\partial u}{\partial \mu}|_0,
\\ \label{18}
\beta_v(u,v) &=& \mu \frac{\partial v}{\partial \mu}|_0,
\\ \label{19}
\gamma_{\phi} &=& \mu \frac{\partial
\ln Z_{\phi}}{\partial \mu}|_0, \\ \label{20}
\gamma_m(u,v) &=& \mu \frac{\partial \ln m}{\partial \mu}|_0 =
\frac{1}{2}\mu \frac{\partial \ln Z_{m^2}^{-1}}{\partial \mu}|_0.
\end{eqnarray}
Using a method of characteristics \cite{difeq} to solve the
partial differential equation (\ref{16}) one can write a formal
solution as:
\begin{equation} \label{21}
\Gamma_R^{N}(m,u,v,\mu) = X(\ell)^{N/2}
\Gamma_R^{N}(Y(\ell)m,u(\ell),v(\ell),\mu \ell),
\end{equation}
where the characteristics are the solutions of the ordinary
differential equations (flow equations):
\begin{eqnarray} \nonumber
\ell \frac{d}{d \ell} \ln X(\ell) = \gamma _{\phi} (u(\ell), v(\ell)),
\qquad
\ell \frac{d}{d \ell} \ln Y(\ell) = \gamma _{m} (u(\ell), v(\ell)),
\\
\ell \frac{d}{d \ell} u(\ell) = \beta _{u} (u(\ell), v(\ell)), \qquad
\ell \frac{d}{d \ell} v(\ell) = \beta _{v} (u(\ell), v(\ell)) \qquad
\qquad \label{22}
\end{eqnarray}
with
\begin{equation} \label{23}
X(1)=Y(1)=1, \qquad u(1)=u, \qquad v(1) = v.
\end{equation}
For small values of $\ell$, equation (\ref{21}) maps the large
length scales (the critical region) to the non-critical point
$\ell=1$.  In this limit the scale-dependent values of the
couplings $u(\ell), \, v(\ell)$ will approach the stable fixed
point, provided such a fixed point exists. The fixed points $u^*,
\, v^*$ of the differential equations (\ref{22}) are given by the
solutions of the system of equations:
\begin{eqnarray} \nonumber
\beta_u(u^*,\, v^*) &=& 0,
\\
\beta_v(u^*, \, v^*) &=&  0.
\label{24}
\end{eqnarray}
The stable fixed point is defined as the fixed point where the
stability matrix
\begin{equation}
B_{ij}= \frac{\partial \beta_{u_i}}{\partial u_j} \, ,
\qquad u_i=\{ u, \, v\}
\label{25}
\end{equation}
possesses eigenvalues $\omega_1, \omega_2$ with positive real
parts. The stable fixed point, which is reached starting from the
initial  values $\ell_0$ in the limit $\ell\rightarrow 0$,
corresponds to the critical point of the system. In the limit
$\ell \rightarrow 0$ (corresponding to the limit of an infinite
correlation length) the renormalized couplings reach their fixed
point values and the critical exponents $\eta$ and $\nu$ are then
given by
\begin{eqnarray} \label{26a}
\eta &=& \gamma_{\phi}(u^*, v^*),
\\
1/\nu &=& 2(1-\gamma_m(u^*, v^*)).
\label{26}
\end{eqnarray}
In the non-asymptotic region  but near the fixed point deviations
from the power laws with the fixed point values of the critical
exponents are governed by the correction-to-scaling exponent
\begin{equation} \label{27}
\omega= {\rm min}(\omega_1, \omega_2)
\end{equation}
in accordance with the Wegner expansion \cite{Wegner72} (\ref{3}).
The rest of the critical exponents are obtained by familiar scaling
laws \cite{Stanley71,rgbooks1}:
\begin{equation}\label{27a}
\alpha=2-d\nu, \mbox{\hspace{3em}}
\beta=\frac{\nu}{2}(d-2+\eta), \mbox{\hspace{3em}}
\gamma=\nu(2-\eta) \,
\end{equation}
 which can be shown to hold from the solutions (\ref{21}).

The  flow equations (\ref{22}) can serve to describe the approach
to  criticality in a larger region where corrections to scaling do
not suffice. As it was mentioned in chapter \ref{I} out of the
asymptotic region physical observables are characterized by
effective exponents introduced to describe a crossover from the
background behaviour to the asymptotic  critical one. In the RG
language they depend on the flow parameter $\ell(\tau)$ through
the dependence of couplings on $\ell$. In particular, according to
the definition (see formula (\ref{2})) for the magnetic
susceptibility effective exponent $\gamma_{eff}$ we have:
\begin{equation} \label{28}
\gamma_{_{eff}}(\tau) = \frac{{\rm d}\ln \chi(\tau)}{{\rm d}\ln \tau} =
\gamma(u(\ell(\tau)), v(\ell(\tau)))+ \dots,
\end{equation}
where the second part is proportional to the $\beta$--functions
and comes from the change of the amplitude part of the
susceptibility. This part is natural to neglect in the vicinity of
a fixed point. Moreover, the contribution of the amplitude
function to the crossover does seem to be small
\cite{Nasser95,Frey90}. Under this restriction, the effective
exponents are simply given by the expressions for the asymptotic
exponents (\ref{26}) but with replacing the fixed point values of
the couplings $u^*, v^*$ by the solutions of the flow equations
(\ref{22}):
\begin{eqnarray} \nonumber
\eta_{_{eff}}(\ell) &=& \gamma_{\phi}(u(\ell), v(\ell)),
\\
\label{29}
1/\nu_{_{eff}}(\ell) &=& 2(1-\gamma_m(u(\ell), v(\ell))).
\end{eqnarray}

In the {\em massive RG scheme} the $Z$-factors
(\ref{11a})--(\ref{14}) are calculated from the vertex functions
(\ref{11}) at zero external momenta $q_1,\dots,q_N$ and non-zero
mass at fixed space dimension in the limit
$\Lambda_0\rightarrow\infty$ . This normalization conditions lead
to the equation for the renormalized vertex functions which is
called Callan--Symanzik equation. Differentiation of the proper
bare vertex function $\Gamma_0^N$ with respect to the renormalized
mass (c.f. eq. (\ref{15})) gives:
\begin{equation} \label{30}
m \frac{\partial}{\partial m} \Gamma_0^{N}|_0=
m \frac{\partial m_0^2}{\partial m}|_0 \Gamma_0^{(1,N)} \, ,
\end{equation}
again, the index $|_0$ means a differentiation at fixed bare parameters and
a new vertex function
\begin{equation} \nonumber
\Gamma_0^{(1,N)} =  \frac{\partial \Gamma_0^N }{\partial m_0^2}
\end{equation}
appears. This vertex function differs from $\Gamma_0^N$ by an
extra term $\varphi^2(R)$ inside the averaging $<\dots>$  in
(\ref{11}). As a result, for the renormalized vertex functions
$\Gamma_R^N$ one gets an inhomogeneous  Callan--Symanzik equation
containing $\Gamma_R^{(1,N)}$ in the right-hand side. However,
close to the critical point $m=0$ the right--hand side can be
neglected with respect to the left--hand side and one arrives to
the homogeneous Callan--Symanzik equation which repeats the
structure of the RG equation (\ref{16}):
\begin{equation} \label{32}
\Big ( m \frac{\partial}{\partial m} +
\beta_u \frac{\partial}{\partial u} +
\beta_v \frac{\partial}{\partial v}
- \frac{N}{2} \gamma_{\phi}\Big)
\Gamma_R^{N}(m,u,v) =0,
\end{equation}
where the coefficients $\beta_u$,  $\beta_u$, $\gamma_\phi$ are
defined by the relations (\ref{17})--(\ref{19}), but the parameter
$\mu$ there is to be understood as the renormalized mass $m$. The
partial differential equation (\ref{32}) is solved again by the
characteristics method sketched by the relations
(\ref{21})--(\ref{23}) and leads to the fixed point relations
given by (\ref{24})--(\ref{27}).

For the sake of completeness let us note that the finiteness of
the renormalized vertex function with one $\phi^2$ insertion
$\Gamma_R^{(1,2)}$ is achieved by the familiar renormalizing
factor $\bar{Z}_{\phi^2}$:
\begin{equation} \label{33}
\Gamma_R^{(1,2)}(k;q,-q;m,u,v) = \bar{Z}_{\phi^2}
\Gamma_0^{(1,2)}(k,q,-q,m_0,u_0,v_0).
\end{equation}
Then, formula (\ref{26}) for the correlation length critical exponent $\nu$
may be recast in terms of $\bar{Z}_{\phi^2}$ by a substitution
$2\gamma_m=\gamma_{\phi}+\bar{\gamma}_{\phi^2}$, which
follows from the relations $Z_{m^2}=\bar{Z}_{\phi^2}Z_{\phi}^{-1}$ and
$\bar{\gamma}_{\phi^2}=\mu {\partial \ln \bar{Z}_{\phi^2}^{-1}}/{\partial
\mu}|_0$. This leads to the relation:
\begin{equation} \label{34}
1/\nu = 2-\gamma_\phi(u^*, v^*) - \bar{\gamma}_{\phi^2}(u^*, v^*).
\end{equation}

The explicit expressions for the RG $\beta$- and $\gamma$-
functions are scheme dependent: they do differ in different
renormalization schemes. Subsequently the fixed point coordinates
are scheme dependent as well. However, the RG functions coincide
in different schemes once calculated at the fixed point. This
leads to the same asymptotic values of the critical exponents:
they are universal and do not depend on renormalization scheme
\cite{rgbooks1,rgbooks2}. In the next subsection we will provide
available expansions for the RG functions of the RIM.

\subsection{Perturbation expansion series and their ``naive" analysis}
\label{V2}

Expressions for the RG functions of the RIM are obtained as series
in the renormalized couplings $u$ and $v$. The perturbation theory
in powers of $u$, $v$ is in fact a perturbation theory in number
of integrations in $k$-space which, on the other side, corresponds
to the number of loops in Feynman diagrams in the diagrammatic
representation for the vertex functions \cite{rgbooks1}. By now,
the RIM RG functions are known up to the order of 5 loops in the
minimal subtraction scheme \cite{Kleinert95} and with record 6
loop accuracy when calculated directly for $d=3$ in the massive
scheme \cite{Carmona00}.

Written in the minimal subtraction scheme the functions read:
\begin{eqnarray} \label{35}
\beta_u &=& - u\,\big(
\varepsilon-u-3/2\,v+17/27\,{u}^{2}+{23}/{12}\,uv+{41}/{32}\,{v}^{2}
+ \dots + \beta_u^{(5LA)} \big),
\\ \label{36}
\beta_v &=& - v\,\big(
\varepsilon -v-2/3\,u+{21}/{32}\,{v}^{2}+{11}/{12}\,vu+{5}/{27}\,{u}^{2},
\\ \label{37}
\gamma_{\phi} &=& 1/54\,u^2 + 1/24\,uv + 1/64\,v^2 + \ldots +
\gamma_{\phi}^{(5LA)},
\\ \label{38}
\bar{\gamma}_{\phi^2} &=& 1/3\,u+1/4\,v -1/9\,u^2 -1/4\,uv -
3/32\,v^2 + \ldots +
\bar{\gamma}_{\phi^2}^{(5LA)}.
\end{eqnarray}
Here, we write down the functions only up to the two--loop approximation
(2LA), referring to the paper \cite{Janssen95} where they were obtained in
3LA. Four- and five--loop contributions may be derived from the RG functions
of the anisotropic cubic model obtained in Ref. \cite{Kleinert95}.

Note, that in the minimal subtraction scheme the dependence on the
space dimension $d$ is trivial and enters expressions
(\ref{35})--(\ref{38}) only via one single term proportional to
$\varepsilon=4-d$ explicitly written in the $\beta$-functions
(\ref{35}), (\ref{36}).  On contrary, in the massive scheme the
space dimension $d$ enters the expressions for the loop integrals,
corresponding to each Feynman diagram of perturbation expansion.
Subsequently the theory is evaluated at space dimension of
interest \cite{loopint}. The series for RG functions of the $d=3$
RIM read:
\begin{eqnarray} \label{39}
\beta_u &=& - u\,\big(1 - u - 3/2\,v + 308/729\,u^2 + 104/81\,uv + 185/216\,v^2
+ \ldots + \beta_u^{(6LA)} \big)\,,
\\ \label{40}
\beta_v &=& - v\,\big(1 - v - 2/3\,u + 95/216\,v^2 + 50/81\,vu + 92/792\,u^2
+ \ldots + \beta_v^{(6LA)} \big)\,,
\\ \label{41}
\gamma_{\phi} &=& 8/729\,u^2 +2/81\,uv + 1/108\,v^2
+ \dots + \gamma_{\phi}^{(6LA)}\,,
\\ \label{42}
\bar{\gamma}_{\phi^2} &=& 1/3\,u + 1/4\,v -2/27\,u^2 -1/6\,uv - 1/16\,v^2
+ \ldots + \bar{\gamma}_{\phi^2}^{(6LA)}\,.
\end{eqnarray}
Again, we display here only the two--loop functions. In the
$\varepsilon$ expansion these functions were first obtained in
Ref. \cite{Grinstein76}. Three--loop terms first reported in Ref.
\cite{Sokolov77} contained some errors, partially corrected in
Ref. \cite{Sokolov81}. Finaly free of errors three--loop
expressions were reported in Ref. \cite{Shpot89}. Subsequently
four--loop series were obtained in Ref. \cite{Mayer89} and only
recently five--loop \cite{Pakhnin00} and record six--loop
\cite{Carmona00} expansions became available.  Universal critical
amplitude  ratios at $d=3$ were first obtained in  three--loop
approximation \cite{Bervillier92} and are known by now  within the
accuracy  of five loops \cite{Mayer98}.

As we noted above, the massive RG scheme does not necessarily mean
evaluation at $d=3$. A possibility to apply the scheme in order to
get the RIM RG functions at arbitrary non-integer space dimension
was outlined in Ref. \cite{Holovatch92}.  For the two--loop RG
functions one gets \cite{Holovatch92}:
\begin{eqnarray}
\beta_u(u,v) & = & -(4-d) u \Big \{ 1 - u - \frac{3}{2} v +
\frac{8}{27} \big [ 9 (i_1-\frac{1}{2})+i_2 \big ] u^2 +
\frac{2}{3} \big [ 12 (i_1-\frac{1}{2})+i_2 \big ] uv +
\frac{1}{4} \big [ 21 (i_1-\frac{1}{2})+i_2 \big ] v^2 \Big \},
\label{43}
\\
\beta_v(u,v) & = & -(4-d) v \Big \{ 1 - v - \frac{2}{3} u +
\frac{1}{4} \big [ 11 (i_1-\frac{1}{2})+i_2 \big ] v^2 +
\frac{2}{3} \big [ 6 (i_1-\frac{1}{2})+i_2 \big ] vu +
\frac{8}{27} \big [ 3 (i_1-\frac{1}{2})+i_2 \big ] u^2 \Big \},
 \label{44}
\nonumber\\ \gamma_{\phi}(u,v) & = & - 2 (4-d) \Big \{ \Big [
\frac{2}{27} u^{2} + \frac{1}{6} uv + \frac{1}{16} v^{2}\Big ] i_2
\Big \}, \label{45}
\\ \label{46}
\bar{\gamma}_{\phi^2}(u,v) & = &  (4-d) \Big \{
\frac{1}{3} u + \frac{1}{4} v -
12 \Big [ \frac{1}{27} u^2 +
\frac{1}{12} uv +
\frac{1}{32} v^2
\Big ] (i_1-\frac{1}{2}) \Big \}.
\nonumber
\end{eqnarray}
The space dimension $d$ enters expressions (\ref{43})--(\ref{46})
also by the $d$-dependent two--loop integrals $i_1(d)$, $i_2(d)$.
Their dependence on $d$ is shown in the figure \ref{figloopint}.
Evaluating the integrals \cite{loopint} for $d=3$: $i_1(3)=2/3$,
$i_2(3)=-2/27$ one recovers the two--loop contributions of the
expressions (\ref{39})--(\ref{42}).  In next order of perturbation
theory for non-integer $d$  the RIM was derived in Ref.
\cite{Holovatch98}, the values of  the corresponding loop
integrals are obtained in Ref. \cite{Holovatch94}.

With the series for the RG functions at hand there are two
different ways to proceed in any of  the two RG approaches
(\ref{35})--(\ref{38}) or (\ref{39})--(\ref{42}). Subsequently,
this leads to four different schemes for  our analysis. Indeed,
the minimal subtraction scheme can be realized by the familiar
$\varepsilon$-expansion \cite{Wilson} as well as directly at $d=3$
\cite{Schloms}. Similarly, the expressions for the RG functions
obtained in the massive scheme can  be directly solved for the
fixed points (FP) or by the pseudo-$\varepsilon$ expansion
\cite{pseps}. However, as we will see below, only two of these
schemes lead to reliable results for the RIM critical behaviour.
Before we proceed let us discuss this situation in more details.

$\bullet$ {\bf 1.} First let us consider the functions
(\ref{35})--(\ref{38}) and perform the $\varepsilon$-expansion. In
order to obtain the expansion, one should (i) solve the fixed
point equations (\ref{24}) with $\varepsilon$ as a small parameter
and then (ii) substitute the coordinates $u^*(\varepsilon)$,
$v^*(\varepsilon)$ as series in $\varepsilon$ into the
RG-$\gamma$-function for the critical exponents for critical
exponents.

Looking for the fixed point solutions in one--loop approximation
(i.e. leaving only linear in $u$, $v$ terms in  the brackets in
(\ref{35}), (\ref{36})) one finds three fixed point (compare with
the fig. \ref{figfp}): the Gaussian fixed point {\bf G}
$u^*=v^*=0$, the pure Ising fixed point {\bf I} $u^*\neq0$,
$v^*=0$ (it describes $n$ non-interacting Ising models: c.f.
Hamiltonian (\ref{8}) with $v=0$) and the ``polymer" fixed point
{\bf P} $u^*=0$, $v^*\neq0$ (putting $u^*=0$ in (\ref{8}) one gets
the $O(n=0)$-model describing  the scaling properties of
self-avoiding walks). It is straightforward to check that the fixed
points {\bf G} and {\bf I} are unstable whereas the fixed point {\bf
P} is stable but as far as $v^*>0$ it is unaccessible for the initial
values of couplings of our model. One does not encounter in 1LA
the fixed point {\bf R} with both non-zero coordinates $u^*\neq0$,
$v^*\neq0$: this is because the system of equations for the fixed
points is degenerate on the one--loop level
\cite{Harris74b,Lubensky75,Grinstein76}. This fixed point appears
in the next, two--loop approximation and leads to the qualitative
picture shown in the figure \ref{figfp}. However, the degeneracy
of the one--loop equations leads to the result, that instead of
expanding in $\varepsilon$ one has to expand in
$\sqrt{\varepsilon}$ \cite{Khmelnitskii75,Grinstein76}. Proceeding
as usual this leads to a  $\sqrt{\varepsilon}$ expansions for the
critical exponents and the stability matrix (\ref{25}) eigenvalues
\cite{Shalaev97,Folk99}:
\begin{eqnarray} \label{47}
\nu  &=& 0.5 +
0.08411582\varepsilon^{1/2}  - 0.01663203\varepsilon  + 0.04775351\varepsilon
^{3/2} + 0.27258431\varepsilon^2\,,
\\ \label{48}
\eta  &=&  -
0.00943396\varepsilon  + 0.03494350\varepsilon ^{3/2} - 0.04486498\varepsilon
^{2} + 0.02157321\varepsilon ^{5/2}\,,
\\ \label{49}
\gamma  &=& 1 +
0.16823164\varepsilon^{1/2} - 0.02854708\varepsilon  + 0.07882881\varepsilon
^{3/2} + 0.56450490\varepsilon ^{2}\,,
\label{sqrteps} \\ \label{50}
\omega_1&=&2\,
\varepsilon+3.704011194 \, \varepsilon^{3/2}+11.30873837 \,\varepsilon^2 \, ,
\\ \label{51}
\omega_2&=&0.6729265850\, \varepsilon^{1/2}-1.925509085 \,
\varepsilon
-0.5725251806 \, \varepsilon^{3/2} -13.93125952 \, \varepsilon^2 \, .
\end{eqnarray}
Two--loop expressions for the exponents were obtained in Ref.
\cite{Grinstein76}, three--loop results were presented
independently in Refs \cite{Jayaprakash77} and \cite{Shalaev77}. The
$\sqrt{\varepsilon}$-expansion series for the RIM amplitudes
ratios \cite{Newlove83} is  also available in the three--loop
accuracy \cite{Shpot90}.  Due to the five--loop results for an
anisotropic cubic model \cite{Kleinert95} it was possible to
obtain four- and five--loop $\sqrt{\varepsilon}$-expansion for the
critical exponents  \cite{Shalaev97} (\ref{47})--(\ref{49}) and
the stability matrix eigenvalues \cite{Folk99}
(\ref{50})--(\ref{51}) of the RIM.

$\bullet$ {\bf 2.} The second way of calculation which may be used
in the minimal subtraction RG approach is  the $d=3$ technique
\cite{Schloms}. It consists in  (i) fixing the value of
$\varepsilon=1$ in (\ref{35}), (\ref{36}), (ii) solving the system
of fixed point equations (\ref{24}) numerically and (iii)
substituting the  numerical values of the fixed point coordinates
into the series for the critical exponents. The RIM
$\beta$-functions are shown in figure \ref{figbetanon} in
two--loop approximation. One can see that they do not possess
non-trivial fixed points at all! However, such behaviour is not
surprising and is not a particular feature of the RIM. It is well
known that perturbation series for the RG functions in the weak
coupling limit are asymptotic at best and should be evaluated by
means of special resummation technique. This will be the subject
of the next Section \ref{VI}, and here, to complete the list of
possible calculation schemes we mention two more schemes in the
massive RG approach:

$\bullet$ {\bf 3.} The massive RG scheme  is implemented by (i)
solving the system of fixed point equations (\ref{24}) numerically
and  then (ii) substituting  the fixed point coordinates numerical
values into the series for the critical exponents. This is the way
which most of the papers are devoted to in the RIM RG analysis
(see the next section for more details).

$\bullet$ {\bf 4.}  The massive RG scheme implemented by a
pseudo-$\varepsilon$ expansion \cite{pseps}. This scheme is based
on the observation that in order to analyze the series
(\ref{39})--(\ref{42}) one can imitate an $\varepsilon$--expansion
introducing an auxiliary parameter $\tau$ instead of the
zeroth--order term $1$ in the massive scheme $\beta$--functions
(\ref{39}), (\ref{40}). Then FPs and critical exponents are
obtained as series in $\tau$ with substitution $\tau=1$ to obtain
their final values. The described pseudo--$\varepsilon$ expansion
\cite{pseps} allows to avoid cumulation of the errors for the
critical exponents from the errors of the fixed point coordinates
and the $\gamma$-- functions. It proved to be highly efficient for
the pure $d=3$ Ising model \cite{LeGuillou80,Guida98}. For the
same reasons as the $\varepsilon$-expansion for RIM turns to the
$\sqrt{\varepsilon}$ expansion, the expansion in $\tau$ turns to a
$\sqrt{\tau}$--expansion. On the basis of the six--loop expansions
(\ref{39})--(\ref{42}) we get the following
$\sqrt{\tau}$--expansion for the RIM  critical exponents and the
stability matrix eigenvalues:
\begin{eqnarray} \label{52}
\nu &=& 0.5+0.102 912 60\,{\tau}^{1/2}+0.012 518 53\,\tau+
0.012 701 78\,{\tau}^{3/2}
+0.056 637 57\,{\tau}^{2}+0.036 943 22\,{\tau}^{5/2}\,,
\\ \label{53}
\eta &=& -0.008 368 20\,\tau+0.021 737 33\,{\tau}^{3/2}-0.014 877 14\,
{\tau}^{2}
+0.017 337 71\,{\tau}^{5/2}\,,
\\ \label{54}
\gamma &=& 1+0.205 825 21\,{\tau}^{1/2}+0.029 221 17\,\tau
+0.015 396 08\,{\tau}^{3/2}
+0.118 581 41\,{\tau}^{2}+0.066 582 80\,{\tau}^{5/2}\,,
\\ \label{55}
\omega_1 &=&
2\,\tau+2.597 611 32\,{\tau}^{3/2}+7.518 005 57\,{\tau}^{2}
+39.868 258 04\,{\tau}^{5/2}\,,
\\ \label{56}
\omega_2 &=&
0.823 300 84\,{\tau}^{1/2}-1.747 132 06\,\tau-1.265 693 50\,{\tau}^{3/2}
-8.750 741 59\,{\tau}^{2}-40.988 383 78\,{\tau}^{5/2}\,.
\nonumber
\end{eqnarray}
These expressions are to be compared with the formulas
(\ref{47})--(\ref{51}).


\section{Series resummation and numerical results}

\label{VI}
\subsection{Summability of RIM divergent series}
\label{VI1}

The core of the perturbation approach described above is the
account of higher perturbation contributions. However, such a way
does not lead necessarily to cumulation of the calculation
accuracy by a ``naive'' summation of successive perturbation
theory terms. It is well known by now that the weak coupling
expansion series for RG functions have zero radii of convergence
and are asymptotic at best \cite{rgbooks2}. Appropriate
resummation  procedures are to be applied in order to handle them.
Early studies of critical phenomena by the $\varepsilon$-expansion
technique lead to a concept of a series ``optimal truncation'' as
a maximal number of terms possessing convergent behaviour. Such
behaviour is typical for asymptotic series \cite{Hardy48} where
series expansion coefficients grow factorially. For example, the
expansion of a RG function $f(u)$ of pure Ising model in powers of
a (single) renormalized coupling $u$:
\begin{equation}\label{57}
f(u) = \sum_k A_k u^k
\end{equation}
was shown to possess the following behaviour
\begin{equation} \label{58}
A_k = c k^{b_0} (-a)^k k! [1 + O(1/k)], \hspace{1em}
k\to\infty
\end{equation}
with known values \cite{Lipatov77,Zinn77,Brezin78} of constants
$c$, $b_0$, $a$. The property (\ref{58})
indicates the Borel summability of the series (\ref{57}). The Borel
resummation procedure takes into account the asymptotic behaviour
of the coefficients (\ref{58}) and maps the asymptotic series (\ref{57})
to a convergent one with the same asymptotic limit. We will describe
the procedure in more details in subsections \ref{VI2} and \ref{VI3}.

It is worthwhile to note, that there does not exists a proof of
Borel-summability  for the $\varepsilon$-expansion of a pure Ising
model. Only expansion in coupling (\ref{57}) is proved to posses
this property \cite{Eckmann75}.  Nonetheless the rich amount of
numerical estimates obtained on the basis of resummed
$\varepsilon$-expansion for the pure Ising model (see e.g.
\cite{rgbooks2}) convinces one in its reliability. However it is
not the case for the $\sqrt{\varepsilon}$-expansion of RIM. The
fact that $\varepsilon$-expansion will not be able to give
information on critical exponents in system with quenched disorder
was predicted already in Refs \cite{Bray87,McKane94} by studying
the randomly diluted model in zero dimensions. The perturbation
theory series of this toy-model appeared to be Borel non-summable.
Moreover, such properties  were shown \cite{McKane94} to be the
direct consequence of the existence of Griffiths-like
singularities \cite{Griffiths69} caused by zeroes of the partition
function of the pure system. Although the
$\sqrt{\varepsilon}$-expansion allowed to predict qualitatively
new critical behaviour of the RIM
\cite{Khmelnitskii75,Grinstein76} it seems to be of no use for a
quantitative analysis. Moreover, from naively adding the
successive perturbational contributions in the
$\sqrt{\varepsilon}$-expansion for the RIM stability matrix
eigenvalues (\ref{50}), (\ref{51}) one observes that already in
the three--loop approximation ($\sim \varepsilon$) $\omega_2$
becomes negative and therefore no stable fixed point exists in
strict $\sqrt{\varepsilon}$-expansion \cite{Folk99}! Even the
resummation procedures applied do not change this picture
\cite{Folk00}. Quite different convergence properties of the
expansions for pure $d=3$ Ising model and RIM may be seen already
applying simple Pad\'e analysis as it was shown in Refs
\cite{Folk98,Folk00}.

The above mentioned divergent properties of the
$\sqrt{\varepsilon}$-expansion concern also  the
pseudo-$\sqrt{\varepsilon}$ expansion (\ref{52})--(\ref{55})
derived in Section \ref{V}. So there remain only two out of four
different ways of numerical analysis of the RIM RG functions. They
are denoted by {\bf $\bullet$ 2} and {\bf $\bullet$ 3} in the
preceding subsection \ref{V2}. Both are based on analysis of RG
series in two couplings. The nature of the RIM RG functions series
in couplings $u$, $v$ remains opened. Nonetheless, the resummation
procedures which in different modifications are used in analysis
of asymptotic series have been applied fruitfully to the RIM RG
series as well
\cite{Jug83,Mayer84,Shpot89,Mayer89,Mayer89a,Holovatch92,%
Janssen95,Folk98,Holovatch98,Folk99,Varnashev00,Folk00,Pakhnin00}.
The method which prevails in the resummation of the RIM RG series
is the generalization of a Pad\'e-Borel resummation technique
\cite{Baker} for the case of two variables. We will review results
 based on such resummation in the next subsection \ref{VI2}.

Recently, it was shown that the expansions for the quenched
diluted Ising model in $d=0$ dimensions are Borel summable,
provided  a specific way of summation is applied \cite{Alvarez00}.
This allowed authors of Ref. \cite{Pelissetto00} both to recover
the asymptotic behaviour of expansion coefficient as well as to
apply to RIM six--loop massive RG functions \cite{Carmona00} the
resummation technique refined by a conformal mapping. This method
will be  reviewed in the last subsection \ref{VI3}.

\subsection{Pad\'e-Borel oriented resummation}
\label{VI2}

In order to perform the Pad\'e-Borel resummation of a truncated
(asymptotic) series in a single variable (e. g. series (\ref{57})
with $L$ terms available) one defines the Borel--Leroy image of
the initial sum $S=\sum_{i=0}^L a_i u^i$ by:
\begin{equation} \label{59}
S^{\rm B}(u)= \sum_{i=0}^L \frac{a_iu^i}{\Gamma(i+q+1)},
\end{equation}
where $\Gamma(x)$ is the Euler's gamma function and $q$ is an
arbitrary non-negative number, which will serve  as a  fit
parameter. The result (\ref{59}) is extrapolated by means of a
rational ap\-pro\-xi\-mant $\left[ M/N \right] (u)$, which is the
quotient of two polynomials in $u$ with $M$ as the order of the
numerator and $N$ as that of the denominator (a Pad\'e approximant
\cite{Baker81}). Subsequently, the resummed function $S^{\rm res}$
is obtained in the form:
\begin{equation} \label{60}
S^{\rm res}(u)=\int_0^\infty {\rm d}t \exp (-t)t^q\left[ M/N
\right] (ut).
\end{equation}
In the single-variable RG functions analysis this procedure was initiated
in Refs \cite{Baker}.

There are different ways to generalize single variable Pad\'e
approximants for the case of two variables $(u,v)$ \cite{Baker81}
and, subsequently to generalize the resummation procedure
(\ref{59}), (\ref{60}). One possibility is to construct the series
in a single dummy variable $x$ (so-called resolvent series
\cite{Watson74}). The variable is introduced by a substitution
$u\rightarrow u x$, $v\rightarrow v x$ and  $x$ must be put equal
to $1$ to obtain final results. In accordance, the Borel-Leroy
image of a truncated series $S=\sum_{0\leq i+j \leq
L}a_{i,j}u^iv^j$ is defined by:
\begin{equation}\label{61}
S^{\rm B}(x) = \sum_{0\leq i+j \leq
L}\frac{a_{i,j}(ux)^i(vx)^j}{\Gamma(i+j+q+1)}.
\end{equation}
Then the series in $x$ is resummed by formula (\ref{60}) and evaluated at
$x=1$. We will refer to this method as to the Pad\'e--Borel method
similarly to the one--coupling case.

Another way to proceed is to make use of the Chisholm approximants
\cite{Chisholm73,Baker81} which are  the generalization of
Pad\'e--approximants to many--variable case. A Chisholm
approximant can be defined as a ratio of two polynomials both in
variables $u$ and $v$, of degree $M$ and $N$ such that the first
terms of its expansion are equal to those of the function which is
approximated. Again, the resummation is performed by (\ref{60})
with Chisholm approximant instead the Pad\'e one. This method will
be referred below as the Chisholm-Borel resummation.

For a given Borel--Leroy image many approximants, both Pad\'e and
Chisholm, can be constructed. However, restrictions are imposed
naturally. Firstly, an approximant should be chosen in the form
re-constituting the sign-alternating high-order behaviour of the
general term of $\beta$- and $\gamma$-functions \cite{Carmona00}.
The approximant generating a sign--alternating series might be
chosen in a form $\left[ M/1 \right]$ with the positive
coefficients at the variable $x$ (or $u$ and $v$). However, these
are the diagonal approximants which give the best convergence in
Pad\'e analysis \cite{Baker81} (see also \cite{Folk00} for a toy
model example). On the other hand, high degree of a polynomial in
the denominator often leads to poles on the positive semi-axis.
One can use analytic continuation and calculate the principal
value of the integral (\ref{60}) to process the singularities
however practical calculations reject the generalization
\cite{Baker}. The reason is both unexpected shift of fixed point
location and different topological structure of the lines of zeros
for the resummed $\beta$-functions. The choice of a Chisholm
approximant is even more complicated since often its coefficients
are undetermined. For instance, already two--variable polynomials
of a second order require two additional conditions to be imposed.
Normally they are chosen to preserve certain symmetry properties.
Provided the couplings $u_0$ and $v_0$ enter the Hamiltonian
(\ref{8}) symmetrically, the approximants must be symmetric in
variables $u$ and $v$ in order not to introduce an additional
symmetry. Another point is that by substitution $v=0$ all the
equations which describe the critical behaviour of the diluted
model are converted into appropriate equations of the pure model.
However, if pure model is solved independently, the resummation
technique with the application of Pad\'e approximant is used.
Thus, Chisholm approximant is to be chosen in such a way that, by
putting any of $u$ or $v$ equal to zero, one reproduces familiar
results for the one-variable case. This also implies a special
choice of additional conditions.

\begin{figure}[htbp]
\epsfxsize=100mm
\epsfysize=70mm
\centerline{\epsffile{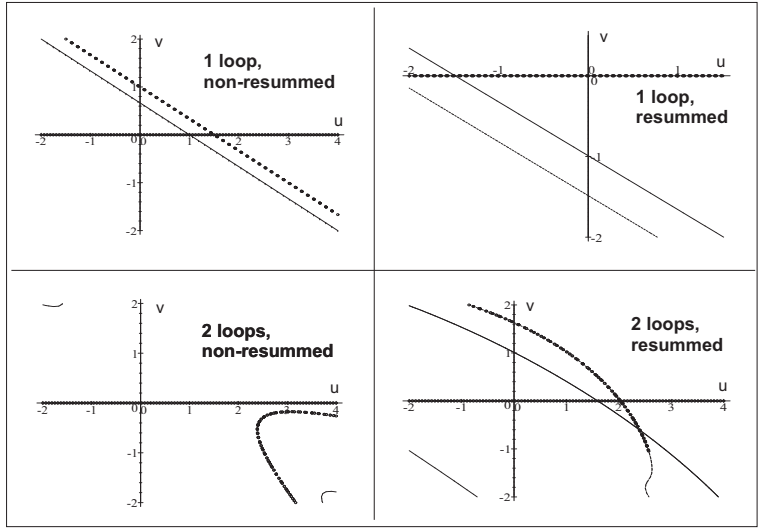}}
\vspace*{5mm}
\caption{
\label{figcomp1} The lines of zeros of non-resummed
(left-hand column) and resummed by the Chi\-sholm-Borel method
(right-hand column) RIM massive $\beta$-functions in different
orders of the perturbation theory: one- and two--loop
approximations. Circles correspond to $\beta_u=0$, thick lines
depict $\beta_v=0$. Thin solid and dashed lines show  zeros of
the analytically continued functions $\beta_u$ and $\beta_v$
respectively. One can see the appearance of the mixed fixed point
$u>0, v<0$ in the two--loop approximation for the resummed
$\beta$-functions. The figure is taken from Ref.
\protect\cite{Holovatch98}.
}
\end{figure}

\begin{figure}[htbp]
\epsfxsize=100mm
\epsfysize=70mm
\centerline{\epsffile{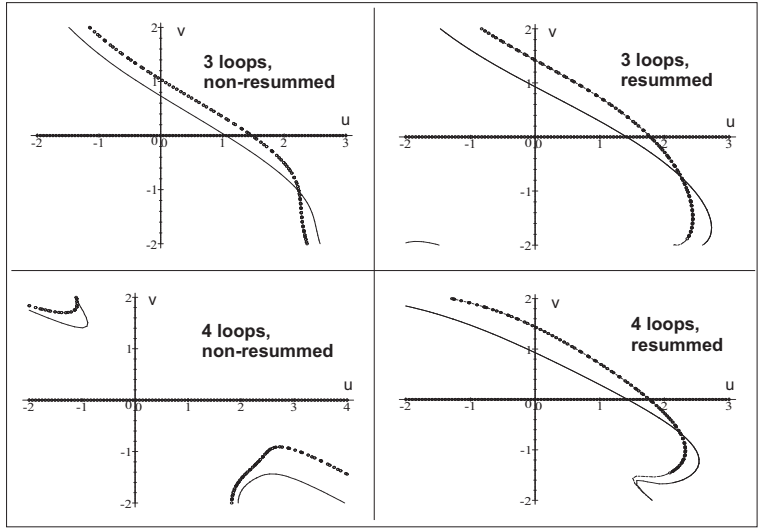}}
\vspace*{5mm}
\caption{
\label{figcomp2} The lines of zeros of non-resummed
(left-hand column) and resummed by the Chi\-sholm-Borel method
(right-hand column)  RIM massive $\beta$-functions in three- and
four--loop approximations. The notations are the same as in the
figure \ref{figcomp1}. Close to the mixed fixed point the
behaviour of the resummed functions remains alike with the
increase of the order of approximation. This is not the case for
non-resummed functions. The figure is taken from Ref.
\protect\cite{Holovatch98}.
}
\end{figure}

Most numerical results on the universal characteristics of RIM at
criticality were obtained on the  basis of the massive
renormalization scheme by numerically solving fixed point
equations (\ref{24}) for the resummed $\beta$-functions
(\ref{39}), (\ref{40}) and resumming the $\gamma$--functions
(\ref{41}),(\ref{42}) in the stable fixed point (scheme {\bf
$\bullet$ 3} in the subsection \ref{V2}). The study of the
massive $\beta$-functions of RIM resummed in this way revealed
that starting from the two--loop approximation the random fixed
point ({\bf R}: see figure \ref{figfp}) is stable and is present
in all orders
of perturbation theory \cite{Jug83,Mayer84,Shpot89,Mayer89,Mayer89a,%
Holovatch92,Holovatch98,Varnashev00,Pakhnin00,Folk00a}. As an
example, in figures \ref{figcomp1}, \ref{figcomp2} we show the
lines of zeros of functions $\beta_u(u,v)$ $\beta_v(u,v)$ in
different orders of perturbation theory without resummation and
resummed by the Chisholm-Borel method. One can see that without
resummation all non-trivial fixed points are obtained only in the
three--loop approximation. Resummation restores presence of
non-trivial fixed points.

The massive scheme of field-theoretical renormalization group was
the basis of first numerical estimates of RIM critical exponents.
Alternatively, the non-perturbative scaling-field approach for
solution of the Wilson's renormalization-group equation was used
to study RIM critical behaviour in Ref. \cite{Newman82}. The
approach similarly to $\varepsilon$-expansion permits to treat
model in continuous dimensions $d$. As a result, RIM critical
exponents were found for $d$, $2.8\leq d\leq 4$. However, the
scaling-field approach was not followed by more precise
calculations and Ref. \cite{Newman82} remains the only theoretical
study were RIM critical exponents were obtained not on the basis
of field-theoretic RG.

\begin{figure}
\epsfxsize=80mm
\centerline{\epsfbox{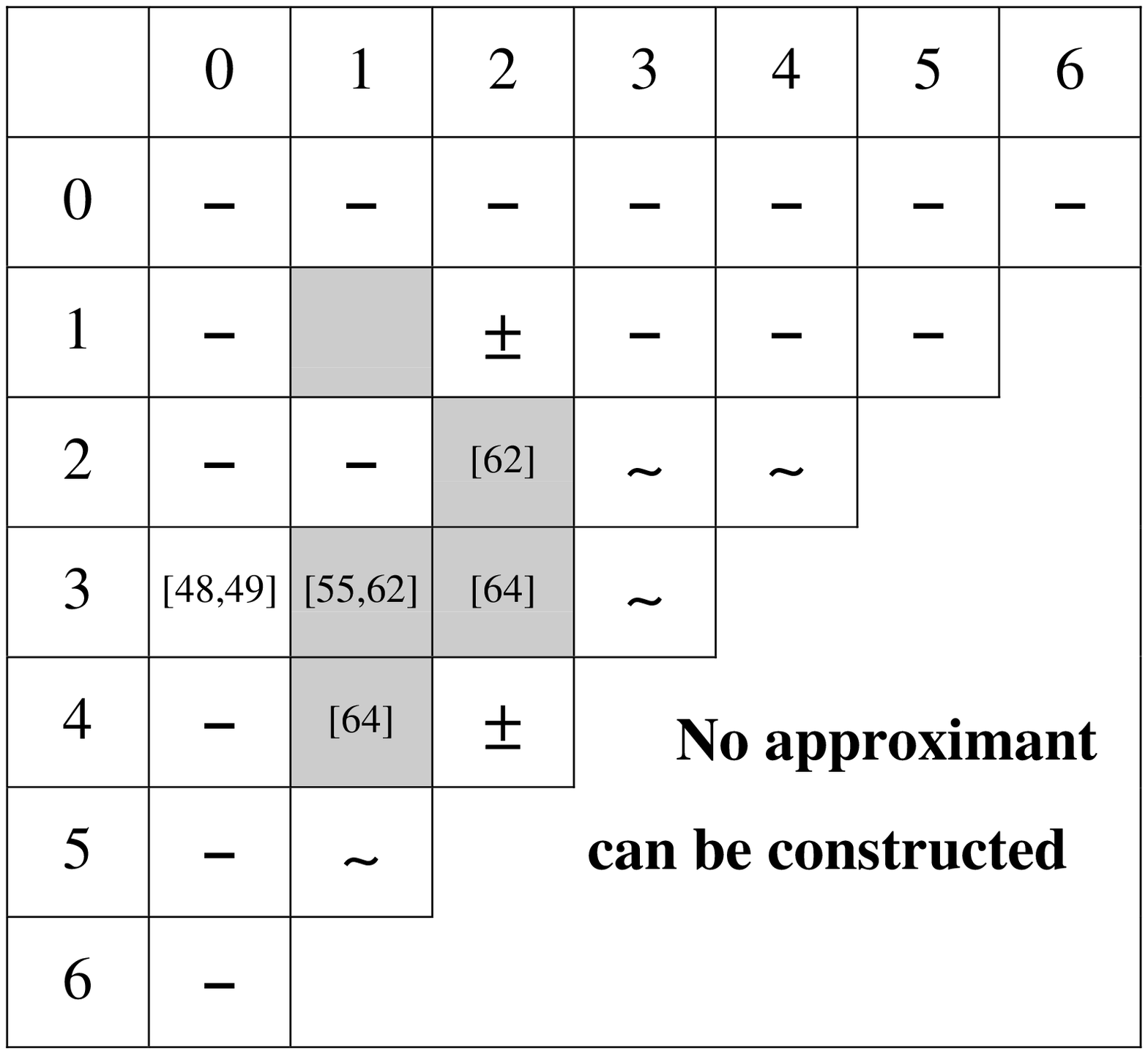}}
\vspace*{5mm}
\caption{\label{table.eps} The fecundity of Pad\'e--Borel
resummation for the RIM $\beta$-functions in massive scheme.
The numbers of a row M and of a column N correspond
to a Pad\'e approximant [M/N] representing (M+N)--loop
approximation. Symbol ``--'' depicts that no fixed point values
are available, ``$\pm$'' and ``$\sim$'' show that a method either
uses analytic continuation due to poles on the real semi-axis or
yields values very far from expected. The fruitful approximants
are shown by dark colour and/or appropriate citations. Though working
approximants are distributed rather stochastically, numerical
results on their basis show evident convergence (table
\ref{table_theory}).}
\end{figure}

Already the study of two--loop RIM massive RG functions resummed by
the Chisholm-Borel procedure \cite{Jug83} revealed that no
difficulties connected to the degeneracy of the $\beta$--functions
are encountered. Critical exponents extracted from resummed
$\gamma$--functions values in the fixed point were found clearly
larger than the pure model ones (see table \ref{table_theory}). As
noted above,  in the three--loop level the straightforward analysis
of the $\beta$--functions \cite{Sokolov77,Sokolov81} yielded fixed point
coordinates and critical exponents without resummation, but the obtained
accuracy did
not allow to estimate, for instance, heat capacity critical
behaviour. On the other hand, the application of the Pad\'e--Borel
technique also encountered hardships within the three--loop level
(see diagram in the Fig.  \ref{table.eps}). Here, the approximant
with linear denominator does not yield a fixed point, while
another near--diagonal approximant $\left[1/2\right]$ is
unreliable because reveals fixed point only when processing the
pole on the real axis via analytic continuation. The same
expressions when treated by means of a Chisholm--Borel technique
allowed to obtain asymptotic critical exponents
\cite{Mayer84,Shpot89}.

 The four--loop results
\cite{Mayer89} were resummed both by means of Chisholm--Borel
\cite{Mayer89}, the first confluent form of the
$\epsilon$-algorithm of Wynn \cite{Mayer89a}
 and Pad\'e--Borel
\cite{Mayer89a,Varnashev00}  methods and showed close results (see
table \ref{table_theory} as well as diagram \ref{table.eps}).
Whereas  Pad\'e--Borel calculations of Ref. \cite{Mayer89a}
exploited the [3/1] Pad\'e approximant, Ref. \cite{Varnashev00}
used a more elaborated generalized Pad\'e--Borel--Leroy
resummation method. The last is  based on exploiting all possible
Pad\'e approximants in the Borel-Leroy resummation
(\ref{59})--(\ref{60}) choosing for each of them the optimal value
of the Leroy parameter $q$ and then averaging the result over all
values given by the approximants \cite{Varnashev00}.

The expressions in the five--loop approximation analyzed with the
application of the fit parameter $q$ required rather artificial
rejecting of many approximants \cite{Pakhnin00}. For instance,
using a criterion that those approximants are working, which
provide maximal stability against variation of $q$, the
approximant $\left[2/2\right]$ was chosen to estimate $u^*$ and
the approximant $\left[3/1\right]$ to obtain $v^*$ . The analysis
allowed the authors of Refs \cite{Pakhnin00} to obtain the five--
loop estimates for the RIM critical exponents (see table
\ref{table_theory}). However, the extend to the six--loop order
revealed a wide gap between five-- and six--loop fixed point
coordinates \cite{Folk00a}  with subsequent inconsistency of the
six--loop values of critical exponents compared with the
five--loop results of Refs \cite{Pakhnin00}. This might serve as
an indirect evidence of a possible non-asymptotic nature of the
series under consideration.

Similar resummation technique was also applied to the minimal
subtraction RG functions (\ref{35})--(\ref{38}) directly at $d=3$
(scheme {\bf $\bullet$ 2} in the subsection \ref{V2})
\cite{Janssen95,Folk98,Folk99,Folk00}. Here, asymptotic as well as
effective critical behaviour was studied. Again, as in the massive
RG case, resummation restores the presence of non-trivial fixed
point in the two--loop approximation (see figure \ref{figbetares})
and preserves it in three \cite{Janssen95} and four loops
\cite{Folk98,Folk00}. However, in the five-loop order the applied
Chisholm-Borel resummation scheme did not lead to a real root for
the random fixed point \cite{Folk98,Folk00}. One of reasons for
such behaviour may be (possible) Borel-non-summability of the
series under consideration. In this case the four--loop
approximation will be an ``optimal truncation'' for the resummed
minimal subtraction perturbation theory series, similar to the
non-resummed asymptotic series, e.g. in the
$\varepsilon$--expansion of $O(n)$--symmetric $\phi^4$ model,
where ``naive'' interpretation of the series truncated by
$\varepsilon^2$ term leads to the best (optimal) result.

\begin{figure}[htbp]
\epsfxsize 100mm
\centerline{\epsffile{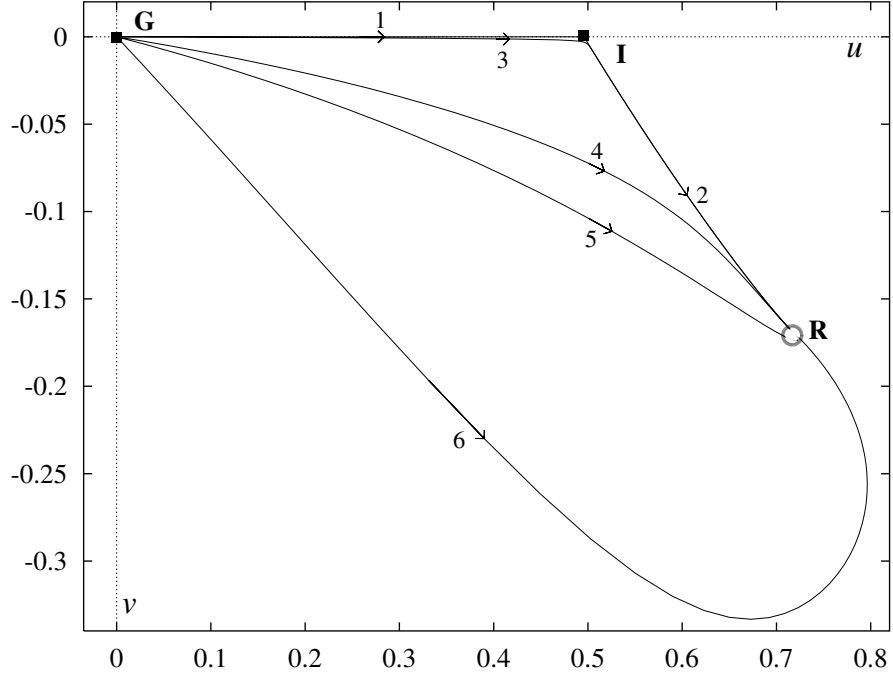}}
\vspace*{5mm}
\caption{\protect\label{flows}  Flow lines for the RIM. Fixed
points {\bf G, I} are unstable, fixed point $R$ is the stable one.
The figure is taken from Ref. \protect\cite{Folk00}.
Note that the authors of Ref. \protect\cite{Folk00} exploited different
in comparison to formulas
(\protect\ref{35})-(\protect\ref{38})
normalization of couplings $u\rightarrow 3\,u$ and $v\rightarrow 8/3\,v$.
}
\end{figure}

\begin{figure}[htbp]
\epsfxsize 80mm
\centerline{\epsffile{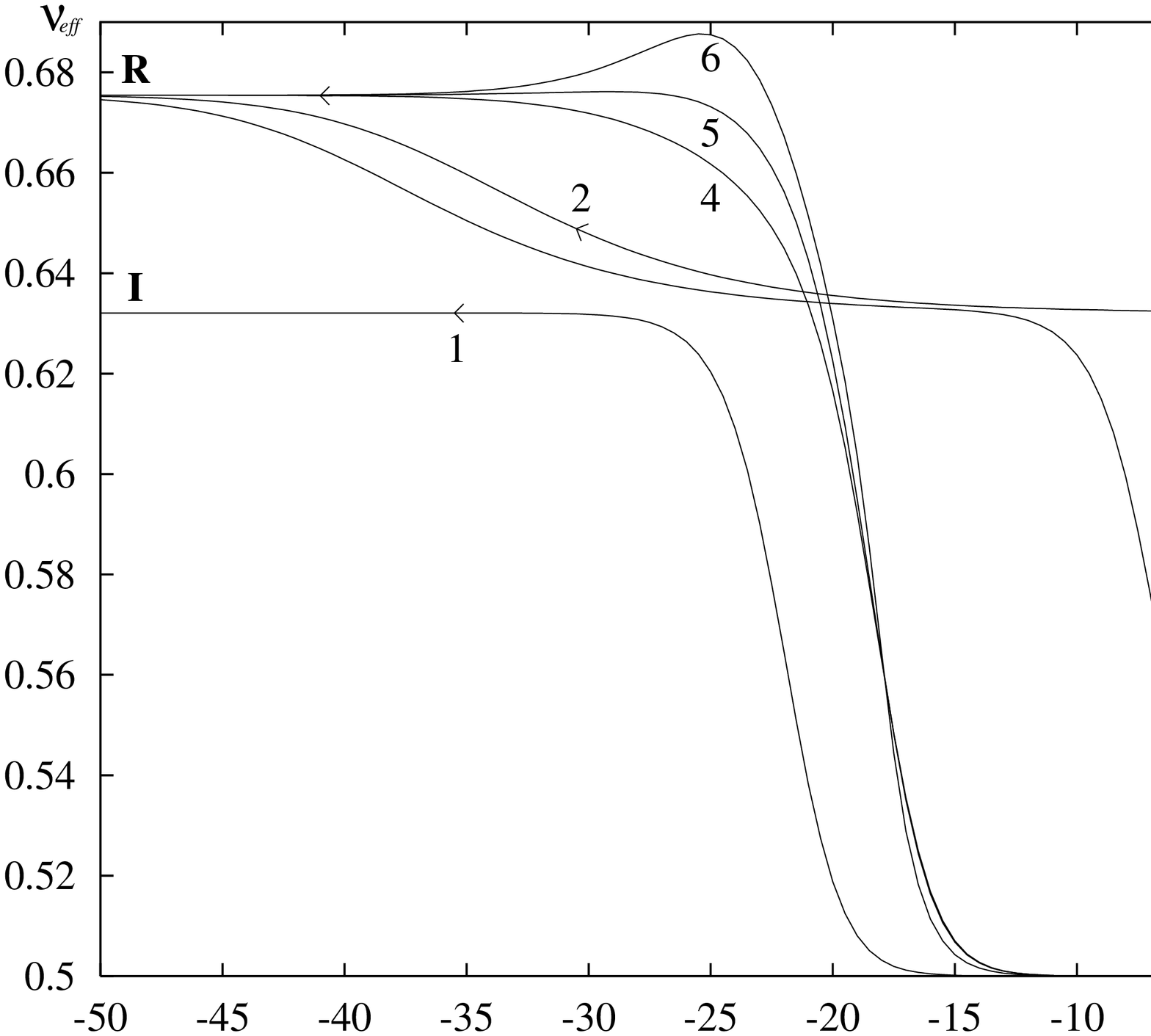}}
\vspace*{5mm}
\caption{\protect\label{nueff} Effective exponent $\nu_{eff}$
versus logarithm of the flow parameter $\ell$ for the flows shown
in Fig. \protect\ref{flows}. The figure is taken from Ref.
\protect\cite{Folk00}. }
\end{figure}

\begin{figure}[htbp]
\epsfxsize 100mm
\centerline{\epsffile{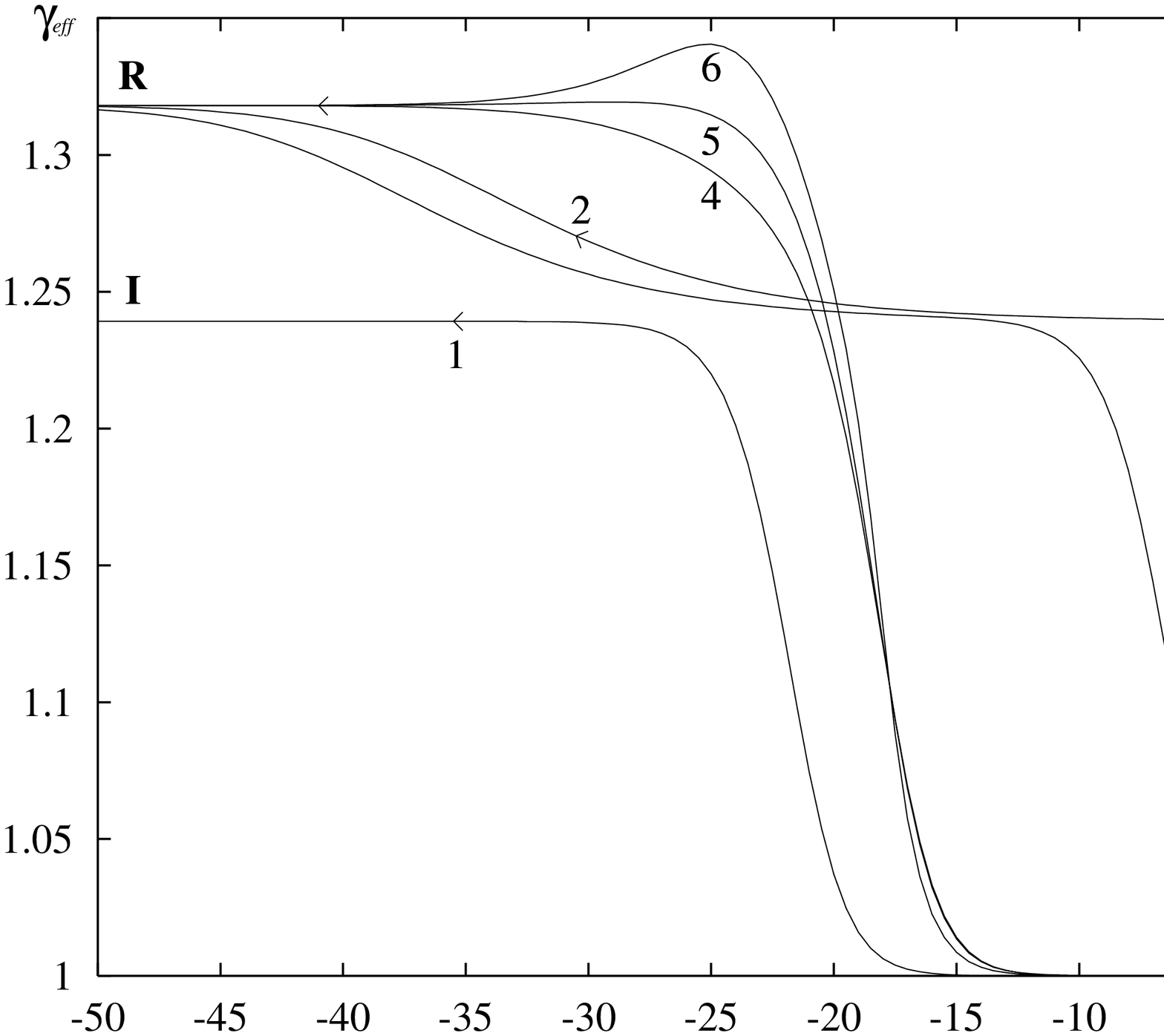}}
\vspace*{5mm}
\caption{\protect\label{gammaeff} Effective exponent
$\gamma_{eff}$ versus logarithm of the flow parameter $\ell$ for
the flows shown in Fig. \protect\ref{flows}. The figure is taken
from Ref. \protect\cite{Folk00}. }
\end{figure}

Obtained in three- \cite{Janssen95} and four-
\cite{Folk98,Folk99,Folk00} loop approximation values of
asymptotic critical exponents are listed in the table
\ref{table_theory}, effective exponents were calculated
\cite{Janssen95,Folk00} by resummation of formulas (\ref{29}),
(\ref{34}). Figure \ref{flows} shows the solutions $u(l)$, $v(l)$
of the flow equations (\ref{22}) calculated by Chisholm-Borel
resummation of four--loop $\beta$-functions (\ref{35}), (\ref{36})
\cite{Folk00}. The flows shown in figure \ref{flows} allow for
several different scenarios for the values of the effective
critical exponents (see figs. \ref{nueff}, \ref{gammaeff}). Both
in experiment as well as in computer simulations (see Tabs.
\ref{table_experiment}, \ref{table_MC}) exponents reported differ
and even exceed the known asymptotic theoretical values. This
non-universal behaviour might be related to the possible
non-asymptotic behaviour found in different flows as has been
suggested in Refs \cite{Heuer93,Janssen95}.  The difference
might be due to (i) the different temperature regions of the
experiment and/or (ii) the different concentrations. The initial
values for the couplings in the flow equations depend on the value
of the concentration, especially for small dilution one expects
the coupling $v$ to be proportional to the concentration. If this
is the case one expects a monotonous increase of the values of the
effective exponent to the asymptotic value. A typical scenario is
seen in curve 3 of Figs.  \ref{flows}, \ref{nueff},
\ref{gammaeff}.  In this case effective exponents equal to the
pure model critical exponents might be found in relatively wide
region of temperature. Then as the attraction region of the fixed
point {\bf I} becomes weaker and weaker, an overshooting is
possible and effective exponents larger their asymptotic values
might be found. This scenario is predicted for larger dilutions
and represented by the curve 6 of Figs. \ref{flows}, \ref{nueff},
\ref{gammaeff}. Curves 5, 6 correspond to situation when crossover
from the mean field behaviour towards the random one is not
influenced by the presence of a pure fixed point  {\bf I}.

\subsection{Resummation based on the conformal mapping}
\label{VI3} The resummation procedure  based on conformal mapping
is widely used in the analysis of asymptotic series for models
with one coupling, in particular for the pure $d=3$ Ising model
(scalar $\phi^4$ theory, see e.g.
\cite{LeGuillou80,Guida98,rgbooks2}). But it assumes that the
behaviour of high order terms of the series expansion is known. It
is not the case for the RIM and hence the Pad\'e-Borel based
resummation technique was used there as described in the former
subsection \ref{VI2}. However, recently by studying the $d=0$
dimensional quenched diluted Ising model it was shown analytically
\cite{Alvarez00} that the perturbative expansions for the free
energy are Borel summable provided the summation is carried in a
special way. The main results of Ref. \cite{Alvarez00} state that
if the double expansion of the RIM functions in powers of $u$, $v$
is written as
\begin{eqnarray} \label{62}
f(u,v)&=&\sum_{i=0}^\infty c_i(v)u^i, \\ \label{63}
c_i(v)&=&\sum_{j=0}^\infty c_{ij}v^j,
\end{eqnarray}
then the expansion (\ref{63}) as well as expansion (\ref{62}) at
$v$ fixed are Borel summable. These results enabled authors of
Ref. \cite{Pelissetto00} to perform the resummation of the
six--loop series (\ref{39})--(\ref{42}) \cite{Carmona00}.
Moreover, by noticing that the large order behaviour of the series
(\ref{63}) can be derived on the basis of known asymptotics
(\ref{58}) \cite{Lipatov77,Zinn77,Brezin78} they were able to
perform the resummation based on conformal mapping for the series
in $v$ (\ref{63}) to get convergent results for the coefficients
$c_i(v)$.

The procedure of conformal mapping resummation of a single-variable
$u$ series is
standard  \cite{LeGuillou80,Zinn81}. For a given Borel-Leroy transform
$S^{\rm B}(u)$ (\ref{59}) the initial series
may be regained from
\begin{equation} \label{64}
S^{\rm res}(u)= \int_{0}^{\infty}{\rm d} t t^b e^{-t} S^{\rm B}(u
t).
\end{equation}
Assuming the behaviour of the high order terms (\ref{58}) one
concludes that the singularity of the transformed series $S^{\rm
B}(u)$ closest to the origin is located at the point $(-1/a)$.
Conformally mapping the $u$ plane onto a disk of radius 1 while
leaving the origin invariant (see figure \ref{figconfmap}),
\begin{equation} \label{65}
 w = \frac{(1+au)^{1/2}-1}{(1+au)^{1/2}+1}, \hspace{3em} u=
\frac{4}{a} \frac{w}{(1-w)^2},
\end{equation}
substituting this into $S^{\rm B}(u)$, and re-expanding in $w$,
one receives a series defined on  the disk with radius 1 in the
$w$ plane. This series is then re-substituted  into Eq.\
(\ref{64}). In order to weaken a possible singularity in the
$w$-plane the corresponding expression is multiplied by
$(1-w)^{\rho}$ and thus one more parameter $\rho$ is
introduced \cite{Zinn81}.

However, the above procedure may be applied only to the RIM RG
series for the coefficients $c_i$ (\ref{63}), as far as the large
order behaviour (\ref{58}) may be derived only for series in $v$
(\ref{63}) \cite{Pelissetto00}. Asymptotics of the resulting
series in $u$ (\ref{62}) is still unknown which causes an
application of the Pad\'e-Borel resummation technique for their
analysis. The last can be applied to the series (\ref{63}) as
well. In summary this leads to different ways of series
resummation: i) applying Pad\'e-Borel method both for series
(\ref{62}) and (\ref{63}) (in Ref. \cite{Pelissetto00} this
way is called a ``double Pad\'e-Borel method") and ii) applying
the conformal method for series (\ref{63}) and the Pad\'e-Borel
method for series (\ref{62}) (``conformal Pad\'e-Borel method").
Both ways were implemented in Ref. \cite{Pelissetto00} taking
special attention to the choice of different Pad\'e approximants
as well as optimizing the results on the basis of fit parameters.
In the table \ref{table_theory} we display the data for the RIM
critical exponents given by the authors on the basis of the
analysis of the six--loop RG functions in such a way that both
estimates obtained by the double Pad\'e-Borel method and the
conformal Pad\'e-Borel method are included (we denote theme as
PdBr-CM). Separately we give the results of the analysis by the
double Pad\'e-Borel method (PdBr--PdBr). Let us note here, that
the conformal mapping technique appears to give more robust
results even resumming  the RG series for fixed $v^*/u^*$.
Applying a procedure of conformal-mapping based resummation of
Ref. \cite{Carmona00} authors of Ref. \cite{Pelissetto00}
obtained for the six--loop RG series values of critical exponents
which are quite in a good agreement with the other estimates (see
table \ref{table_theory}). However the estimates of the fixed
point coordinates differ essentially in different calculation
schemes. It leads to the conclusion \cite{Pelissetto00} that
probably the optimal truncation of the RIM $\beta$-functions
corresponds to a shorter series than a six--loop one.


\section{Conclusions}

\label{VII} In the present article we have reviewed results
obtained so far in the description of critical properties of a
three dimensional weakly diluted quenched Ising model (RIM).
Following the prevailing bulk of experimental, Monte-Carlo, and
theoretical studies we focused our attention on critical exponents
of the model. It can be seen from the number of relevant papers
that the precise determination of  the values of the exponents was
a challenge justifying the scientific effort. The reason for this
is two-fold.  (i) The RIM allows to include in a simple model the
macroscopic effects of disorder always present in real substances.
(ii) The study of the influence of disorder on universal
properties of critical behaviour besides practical needs is of
high academic interest. It is the domain very close to critical
temperature where even  a very small amount of impurities can
change the properties drastically in comparison with an ideal
magnet. In accordance with the heuristic Harris criterion
\cite{Harris74}, only when the ideal system specific heat is
divergent at criticality, the disordered magnet is characterized
by new critical exponents. The change of critical exponents of RIM
is well established in experimental as well as Monte-Carlo and
theoretical studies. However,  the  numerical values obtained show
much worse self-consistency  compared with the  situation in
studies of pure Ising magnets. This happens due to both technical
and principal difficulties. Moreover, often  the technical
difficulties in determination of the RIM exponents are related to
the principal ones!

Though it is commonly believed that various defects are inevitable
in experimental samples, the number of experimental RIM studies is
much less than those devoted to the determination of the Ising
model critical exponents. As  an explanation one might mention
that the Ising model universality class includes not only magnets
but also simple fluids, ferroelectrics, binary alloys, etc.
Moreover the RIM itself can  show an effective critical behaviour
of the same type as the behaviour in the Ising model universality
class (!). The data for experimentally determined RIM critical
exponents are collected in table \ref{table_experiment}. As  an
peculiarity it is interesting to note that already in early
studies \cite{Dunlap81} critical behaviour different from the pure
Ising model has been observed. Surprisingly enough, since this
work neither the experimental precision of the determination of
the critical exponents  has increased nor a narrower temperature
interval around the critical point could be accessed. This can be
explained by noting that starting from the end of eighties the
researchers' attention shifted to random-field Ising model and the
data for RIM were obtained only as  side product results. It is
worthwhile to note that the theoretically predicted critical
exponents for quenched disorder \cite{Pelissetto00}:
$\nu\simeq0.678, \gamma\simeq1.330$ and for annealed disorder
\cite{note4}: $\nu \simeq0.708, \gamma\simeq1.391$,  are different
but only by a small amount. Most   probably, in real samples one
encounters intermediate situations and exact coincidence between
experiments on diluted crystalline antiferromagnets and
theoretical calculations is hardly expected.

While the existence of a new universality class for the RIM has
been observed already in early experimental studies, this was not
the case in Monte-Carlo simulations. Due to the lack of a proper
finite size analysis a continuous dependence of the critical
exponents on concentration was observed. The exponents were
interpreted as effective critical exponents
\cite{Kouvel64,Riedel74}. In the following it was crucial to
recognize the role of the correction-to-scaling exponent in the
analysis of the simulation data
\cite{Parisi98,Wiseman98a,Wiseman98b}. This allowed to conclude
from the numerical data on the concentration-independent
asymptotic critical exponents \cite{Parisi98}. As one can note
from the table \ref{table_MC} the improvement of computers as well
as of the calculation algorithms allowed to increase the accuracy
of critical exponents. Recently  in Monte-Carlo simulations of the
RIM attention was paid also to study  of the problems of
self-averaging in diluted systems \cite{Marques00,Marques00a}.

The results of experimental studies and Monte-Carlo simulations
are confirmed  by theoretical calculations. The numerical values
of critical exponents obtained by different theoretical methods
are collected in the table \ref{table_theory}. Note, that all
theoretical results were obtained within renormalization-group
approach, and most of them within its field-theoretical formalism
on the basis of the effective Landau-Ginzburg-Wilson Hamiltonian
(\ref{8}). Though the last approach encounters intrinsic obstacles
such as problems with the breakdown of replica symmetry and the
possible existence of Griffiths phase \cite{Griffiths69}, it
remains the only method which can calculate the asymptotic values
of exponents. For the pure Ising model many results are also based
on other methods such as high- and low-temperature expansions.

Another evident difference of the RIM in comparison with  the
Ising model lies in the applicability of different regularisation
schemes of field-theoretical renormalization-group approach. In
the case of Ising model both massive and minimal subtraction
scheme followed by $\varepsilon$-expansion provide consistent and
reliable results. For the RIM, the $\varepsilon$-expansion
degenerates into $\sqrt\varepsilon$-expansion
\cite{Khmelnitskii75,Grinstein76,Shalaev97,Folk99} which seems to
be of no use for quantitative studies \cite{Folk98,Folk00}.
Moreover, the initial series in the couplings, both in the massive
and the minimal subtraction schemes, seem not to be Borel summable
\cite{Bray87,McKane94,Alvarez00}. Nevertheless, Pad\'e-Borel-like
resummation procedures have been employed and allowed to obtain
convergent sequences for the critical exponents from low order
series. This is evident from the agreement of the theoretical
results obtained in different orders of perturbation theory (table
\ref{table_theory}) with the data of experimental (table
\ref{table_experiment}) and Monte-Carlo (table \ref{table_MC})
studies. However, the resummation failed in higher orders which
resulted in the conjecture that there exists an optimal truncation
for the RIM RG functions \cite{Folk00}.

On the  basis of the six--loop RG functions of  the massive
schemes result could only be obtained \cite{Pelissetto00} by means
of  a highly sophisticated resummation procedure \cite{Alvarez00}.
In the frames of this scheme, the estimates for critical exponents
are characterized by the same order of accuracy as for the Ising
model, which are based \cite{Guida98} on six--loop expansions for
RG $\beta$-functions and seven--loop for $\gamma$-functions.
However, the determination of error-bars in the theoretical
calculations is a difficult problem which is solved in various
ways \cite{Varnashev00,Pelissetto00}. Here, the error-bars seem to
measure rather uncertainty of specific theoretical procedures than
the confidence intervals of true values.

For a comparison of the theoretical predictions with experimental
and simulation data one should use effective critical exponents
which have been calculated for the RIM within the RG approach
\cite{Janssen95,Folk00}.

At the very end of this review we want to attract attention again
to the fact that the new critical behaviour corresponding to the
RIM universality class was experimentally observed so far only in
magnetic  systems like (anti)ferromagnets (see Table
\ref{table_experiment}). Still it remains a challenge to set up an
experiment showing RIM-behaviour in other condensed matter
systems. A  promising example might be a fluid near its
liquid--gas critical point  in a porous medium. Recently, precise
experiments on the critical behaviour of liquid helium-4  near the
superfluid transition in porous medium \cite{He4dil} confirmed the
irrelevancy of quenched disorder. This was expected from the
Harris criterion \cite{note5} since the specific heat exponent
near the superfluid transition is roughly zero. Thus an
experimental study of a simple liquid in  a porous medium may
provide the first observation of a  RIM critical behaviour in a
non-magnetic system \cite{note6}.


\section*{Acknowledgements}
\label{VIII}

Writing this review and performing studies of RIM we benefited
from discussions and contacts with numerous colleagues. We greatly
acknowledge useful discussions with: David Belanger, Bertrand
Berche, Christian von Ferber, Hagen Kleinert, Victor
Mart\'in-Mayor, Alan McKane, Verena Schulte-Frohlinde, Mykola
Shpot, Aleksandr Sokolov. The materials presented here were used
in part during lectures given by one of us in summer semester 2001
in the Universit\'e Henri Poincar\'e, Nancy 1 (France). Yu.H
expresses his thanks to Bertrand Berche and colleagues at the
Laboratoire de physique des mat\'eriaux (Universit\'e Nancy 1) for
their hospitality, support and many useful comments.

Work was supported in part by ``\"Osterreichische Nationalbank
Jubil\"aumsfonds'' through the grant No 7694.

\newpage


\section{Tables and Figures}

{\footnotesize
\begin{table}
\caption{\label{table_experiment} {The experimentally measured
critical exponents of the materials, which correspond to random
Ising models. The measurement procedures are given in the
following notations: NMR --- nuclear magnetic resonance; LB
--- linear birefringence; NS --- neutron scattering; MS ---
M\"ossbauer spectroscopy; SMXS --- synchrotron magnetic x-ray
scattering; XS --- x-ray scattering; $\tau$ denotes the reduced
temperature interval, where the power--law fit was done or the
minimal value of the reduced temperature reached in an experiment.
} }
\begin{center}
\tabcolsep0.2mm
\begin{tabular}{|c|c|c|c|c|c|c|c|}
Ref. & material & method &$|\tau|$&
$\beta$ & $\gamma$& $\nu$ & $\alpha$ \\
\hline
Dunlap et al., 1981
& ${\rm Mn_pZn_{1-p}F_2}$& NMR&$10^{-3}$&
                                        0.349$\pm$0.008
                                             & --- & ---&
\\ Ref. \cite{Dunlap81}
&            p=0.864           &&&&&&
\\
\hline
Birgeneau et al., 1983
&
             ${\rm Fe_pZn_{1-p}F_2}$&
                         NS; LB&$10^{-1} \div 2 \cdot 10^{-3}$&
                                   --- &
                                            1.44$\pm$0.06&
                                                       0.73$\pm$0.03&
$-0.09\pm 0.03$
\\ Ref. \cite{Birgeneau83}
&           p=0.6, 0.5 && $2 \cdot 10^{-2} \div 2 \cdot 10^{-3}$  &&&&
\\
\hline
Hastings et al., 1985
&
                ${\rm Dy_3 Al_5 O_{12}}$&
                         NS&$4 \cdot 10^{-2}$&
                                 0.350$\pm$0.01&
                                                 --- & 0.73 &
\\ Ref. \cite{Hastings85}
&             +1\% Y powder &&&&&&
\\
\hline
Belanger et al., 1986
&
         ${\rm Fe_pZn_{1-p}F_2}$&
                           NS& $10^{-1} \div 1.5 \cdot 10^{-3}$&
                                   --- & 1.31$\pm$0.03&
                                                 0.69$\pm$0.01&
\\ Ref. \cite{Belanger86}
&              $p=0.46$ &&&&&&
\\
\hline
Barret, 1986
&
                ${\rm Fe_pZn_{1-p}F_2}$&
                            MS&$(4.7\div36\cdot 10^{-3})\div 1 \cdot 10^{-3}$&
                                    0.36$\pm$0.01&
                                                      --- & ---&
\\ Ref. \cite{Barret86}
&               $p=0.9925\div0.95$ &&&&&&
\\
\hline Mitchell et al., 1986 & ${\rm Mn_pZn_{1-p}F_2}$& NS& &&&&\\
Ref. \cite{Mitchell86} &$p=0.75$&&$2\cdot10^{-1} \div 4\cdot
10^{-4}$&--- & 1.364$\pm$0.076&
0.715$\pm$0.035&
\\
&$p=0.5$&&$1\cdot 10^{-1}\div5\cdot 10^{-3}$, $\tau>0$&--- & 1.57$\pm$0.16&
0.75$\pm$0.05&
\\
&$p=0.5$&& $1\cdot 10^{-1}\div5\cdot 10^{-3}$, $\tau<0$&--- & 1.56$\pm$0.16&
0.76$\pm$0.08&
\\
\hline
Thurston et al., 1988&
             ${\rm Mn_pZn_{1-p}F_2}$&
                               SMXS&$6 \cdot 10^{-2}\div 1 \cdot 10^{-3}$&
                                      0.33$\pm$0.02&
                                                       --- & ---&
\\ Ref. \cite{Thurston88}
&                 $p=0.5$ &&&&&&
\\
\hline
Rosov et al., 1988&
                 ${\rm Fe_pZn_{1-p}F_2}$&
                              MS&$1 \cdot 10^{-1}\div 3 \cdot 10^{-4}$&
                                      0.350$\pm$0.009&
                                                       --- & ---&
\\ Ref. \cite{Rosov88}
&                $p=0.9$ &&&&&&
\\
\hline
Ramos et al., 1988&
                 ${\rm Mn_pZn_{1-p}F_2}$&
                              LB& $10^{-2}$? &   --- &
                                 --- & ---& -0.09$\pm$0.03
\\ Ref. \cite{Ramos88}
&                $p=0.40, 0.55, 0.83$ &&&&&&
 \\
\hline
Ferreira et al., 1991&
                 ${\rm Fe_pZn_{1-p}F_2}$&
                              LB& ? & &
                                 --- & --- & -0.09
\\ Ref. \cite{Ferreira91}
&                $0.31\leq p \leq 0.84$ &&&&&&
  \\
\hline
Belanger et al., 1995 &
                 ${\rm Fe_pZn_{1-p}F_2}$&
                              NS&  $<10^{-1}$& 0.35 &
                                 --- & ---& ---
\\  Ref. \cite{Belanger95}
&                $p =0.5$ &&&&&&
  \\
\hline
Belanger et al., 1996 &
                 ${\rm Fe_pZn_{1-p}F_2}$&
                              NS& $10^{-2}$&  0.35 &
                                 --- & ---& ---
\\ Ref. \cite{Belanger96}
&                $p =0.52$ &&&&&&
  \\
\hline
Hill et al., 1997 &
                 ${\rm Fe_pZn_{1-p}F_2}$&
                              XS&  $10^{-2}$&  0.36$\pm$ 0.02&
                                 --- & ---& ---
\\ Ref. \cite{Hill97}
&                $p =0.5$ &&&&&&
\\
\hline
Slanic et al., 1998&
                 ${\rm Fe_pZn_{1-p}F_2}$&
                              LB&  ? & --- &
                                 --- & ---& $-0.10\pm0.02$
\\ Ref. \cite{Slanic98a}
&                $p =0.93$ &&&&&&
  \\
\hline
Slanic et al., 1998&
                 ${\rm Fe_pZn_{1-p}F_2}$&
                              NS&   ? & ---&
                                 1.35$\pm$0.01 & 0.71$\pm$0.01& ---
\\ Ref. \cite{Slanic98} & $p =0.93$ &&&&&&
  \\
\hline
Slanic et al., 1999&
                 ${\rm Fe_pZn_{1-p}F_2}$&
                              NS& $10^{-2}\div1.14\cdot 10^{-4}$&  ---&
                                 1.34$\pm$0.06 & 0.70$\pm$0.02& ---
\\ Ref. \cite{Slanic99}
&                $p =0.93$ &&&&&&
\\
\end{tabular}
\end{center}
\end{table}
}

{\footnotesize
\begin{table}
\caption{\label{table_MC}
{The critical exponents of the RIM as they are
obtained in MC simulations. The asterisk at concentration denotes that
disorder was realized in a canonical manner.}
}
\begin{center}
\tabcolsep0.6mm
\begin{tabular}{|c|c|c|c|c|c|c|}
Ref.     & max. size& concentr. range & $p$ & $\beta$ & $\gamma$ &
$\nu$
\\
\hline
Landau, 1980 &  30  &  $0.4 < p \leq 1$   & {\bf
all} &  $0.31$ & $1.25$   & --- \\
Ref. \cite{Landau80} &&&&&&
\\
\hline
                 &      &                     & 1  & $0.30 \pm 0.02$ &---
& ---
\\
                 &      &                 & 0.985  & $0.31 \pm 0.02$ &---
& ---
\\
Marro et al., 1986
&  40 & $0.8 \leq p \le 1$  & 0.95   &
$0.32 \pm 0.03$ &---& ---
\\ Ref. \cite{Marro86}
                      &      &                     & 0.9    & $0.355 \pm 0.010$
&---&---
\\
                      &      &                     & 0.8    & $0.385 \pm 0.015$
&---&---
\\
\hline
                      &      &                     & 1     & $0.29 \pm 0.02$
&---&---
\\
Chowdhury et al., 1986
& 90   & $0.8 \leq p \le 1$  & 0.95   &
$0.28 \pm 0.02$ &---&---
\\  Ref. \cite{Chowdhury86}
                      &      &                     & 0.90   & $0.31 \pm 0.02$
&---&---
\\
                      &      &                     & 0.80   & $0.37 \pm 0.02$
&---&---
\\
\hline
Braun et al., 1988
  &   40& ---                 & 0.80   & $0.392 \pm 0.03$ &---&---
\\ Ref. \cite{Braun88} &&&&&&
\\
\hline
Wang et al., 1989
&  100  & $0.4 \leq p \leq 0.8$   &
{\bf all}    &---&
 $1.52 \pm 0.07$ & $0.77 \pm 0.04$
\\  Ref. \cite{Wang89} &&&&&& \\
\hline
Wang et al., 1990
   & 300   & ---                  & 0.8   &---&$1.36\pm
0.04$&---
\\ Ref. \cite{Wang90} &&&&&& \\
\hline
                      &       &                    & 1     & --- & --- &
0.629(4)
\\
Holey et al., 1990 &  64    &  $0.8 < p \leq 1$  & 0.9   & ---
& --- & $ < 2/3$
\\
Ref. \cite{Holey90}  &       &                    & 0.8   & --- & --- &
0.688(13)
\\
\hline
                      &       &                    & 1
& $0.305\pm0.01$ & $1.24\pm0.01$&---
\\
                      &       &                   & 0.9   & $0.315\pm0.01$
& $1.30\pm0.01$&---
\\
Heuer, 1990 &  60   & $0.5 < p \leq 1$   & 0.8   &
$0.330\pm0.01$ & $1.35\pm0.01$&---
\\
Ref. \cite{Heuer90} &      &                    & 0.6   & $0.330\pm0.01$
& $1.48\pm0.02$&---
\\
                      &      &                    & 0.5   & $0.335\pm0.01$
& $1.49\pm0.02$&---
\\
\hline
                      &      &                    & 1     & $0.33\pm 0.01$
& $1.22\pm0.02$ & $0.624 \pm 0.010$
\\
                      &      &                    & 0.95  & $0.31\pm0.02$
& $1.28\pm0.03$ & $0.64 \pm 0.02$
\\
Heuer, 1993 &  60  & $0.6 < p \leq 1$   & 0.9   &
$0.31\pm0.02$ & $1.31\pm0.03$ & $0.65 \pm 0.02$
\\ Ref. \cite{Heuer93} &      &                    & 0.8   & $0.35\pm0.02$
& $1.35\pm0.03$ & $0.68 \pm 0.02$
\\
                      &      &                    & 0.6   & $0.33\pm0.02$
& $1.51\pm0.03$ & $0.72 \pm 0.02$
\\
\hline
Hennecke et al., 1993
&  90   & ---             & 0.6 & $0.42 \pm 0.04$
&  &$0.78 \pm 0.01$
\\ Ref. \cite{Hennecke93} &&&&&& \\
\hline
Wiseman et al., 1998
 & 64  & ---               &  0.8 & $0.344 \pm 0.003$
& $1.357 \pm 0.008$ & $0.682 \pm 0.003$
\\ Ref. \cite{Wiseman98a} &&&&&& \\
\hline
Wiseman et al., 1998
& 90  & ---               &  0.6$^*$ & $0.316 \pm 0.013$
& $1.522 \pm 0.031$ & $0.722 \pm 0.008$
\\ Ref. \cite{Wiseman98b}
                        & 80  & ---               &  0.6 & $0.313 \pm
0.012$
& $1.508 \pm 0.028$ & $0.717 \pm 0.007$
\\
\hline
Ballesteros et al., 1998
& 128 & $0.4 \leq p \leq 0.9$ & {\bf
all} & $0.3546 \pm 0.0028$ & $1.342 \pm 0.010$ & $0.6837 \pm
0.0053$
\\ Ref. \cite{Parisi98} &&&&&& \\
\hline
Marques et al., 2000
& 60 & $0.8 \leq p \leq 0.9975$ & {\bf
all} & $0.3546 $ & $1.342 $ &
\\ Ref. \cite{Marques00a} &&&&&& \\
\hline
Marques et al., 2000
& 100 & $0.5$ & & $ $ & & 0.6837
\\ Ref. \cite{Marques00b} &&&&&& \\
\end{tabular}
\end{center}
\end{table}
}

{\footnotesize
\begin{table}
\caption{\label{table_theory}
The theoretical values for asymptotic critical exponents of
the RIM. {\it n}LA denotes the {\it n}th order in loopwise approximation
within massive (`mass') and $3d$ minimal subtraction (`MS') schemes
of field-theoretical renormalization group approach.
The resummation procedures are given in the following notations:
ChBr -- Chisholm--Borel; PdBr -- Pad\'e--Borel;
AW -- $\varepsilon$ algorithm of Wynn,
CM -- Borel transformation with conformal mapping.
SF stands for Golner-Riedel scaling field approach,
superscript $^c$ at the correction-to-scaling
exponent $\omega$ indicates that
the real part of corresponding complex number
is shown.
}
\tabcolsep0.6mm
\begin{tabular}{|c|c|c|c|c|c|c|c|}
Ref. & RG scheme &
Order & Resummation & $\nu$ & $\eta$ & $\gamma$ & $\omega$
\\
\hline
Sokolov et al., 1981 & mass & 3LA & No & & 0.009 & 1.31 &
\\ Ref. \cite{Sokolov81} &&&&&&& \\
\hline Newman et al., 1982 & SF & No &  & 0.70 & 0.015 & 1.39 &
0.41
\\  Ref. \cite{Newman82} &&&&&&& \\
\hline
Jug, 1983 & mass & 2LA & ChBr & 0.678 & 0.031 & 1.336 & 0.450$^{c}$
\\  Ref. \cite{Jug83} &&&&&&& \\
\hline
Mayer et al., 1984  & mass & 2LA & ChBr & & 0.031 & 1.337 &
\\
Ref.  \cite{Mayer84} &      & 3LA & ChBr & & 0.022 & 1.325 &
\\
\hline
Mayer et al., 1989 & mass & 4LA & ChBr & 0.670 & 0.034 & 1.326 &
\\  Ref. \cite{Mayer89} &&&&&&& \\
\hline
Mayer, 1989 & mass & 4LA & AW & 0.6680 & & 1.318 &
\\
Ref. \cite{Mayer89a} &      & 4LA & PdBr & 0.6714 & & 1.321 &
\\
\hline
Shpot, 1989 & mass & 3LA & ChBr & 0.671 & 0.021 & 1.328 & 0.359
\\  Ref. \cite{Shpot89} &&&&&&& \\
\hline
Janssen et al., 1995 & MS, 3d & 3LA & PdBr & 0.666 & & 1.313 & 0.366
\\  Ref. \cite{Janssen95} &&&&&&& \\
\hline
Holovatch et al, 1997 & mass & 3LA & ChBr & 0.671 & 0.019 & 1.328 & 0.376
\\  Ref. \cite{Holovatch98} &&&&&&& \\
\hline
    &        & 2LA &           & 0.665 & 0.032 & 1.308 &
0.162
\\
Folk et al., 1998 & MS, 3d & 3LA & ChBr  & 0.654 & 0.022 & 1.293 &
0.430
\\
Ref. \cite{Folk98}  &        &  4LA  &  & 0.675 & 0.049 & 1.318 &
0.390$^c$
\\
\hline Folk et al., 1999 & MS, 3d & 4LA & ChBr & & & &
0.39(4)$^{c}$
\\  Ref. \cite{Folk99}  & mass & 4LA & ChBr & & & & 0.372(5)
\\
\hline
Pakhnin et al., 2000 & mass & 5LA & PdBr & 0.671(5) & 0.025(10)&
1.325(3) &
0.32(6)
\\  Ref. \cite{Pakhnin00} &&&&&&& \\ \hline
Varnashev, 2000 & mass & 4LA & PdBr & 0.681(12) & 0.040(11) & 1.336(20) &
\\
Ref. \cite{Varnashev00}  &      &     & PdBr & 0.672(4) & 0.034(10)
& 1.323(10) &
0.330
\\
\hline
Pelissetto et al., 2000 & mass & 6LA & PdBr-CM & 0.678(10) & 0.030(3) &
1.330(17) &
0.25(10) \\  Ref. \cite{Pelissetto00} & & & PdBr-PdBr& 0.668(6)  &
0.0327(19)  &  1.313(14)  & 0.25(10)\\
\end{tabular}
\end{table}
}


\newpage
\vspace*{10mm}

\vspace*{15mm}

\newpage
\vspace*{10mm}

\newpage
\vspace*{10mm}

\newpage
\vspace*{10mm}
\begin{figure}[htbp]
\epsfxsize 100mm
\centerline{\epsffile{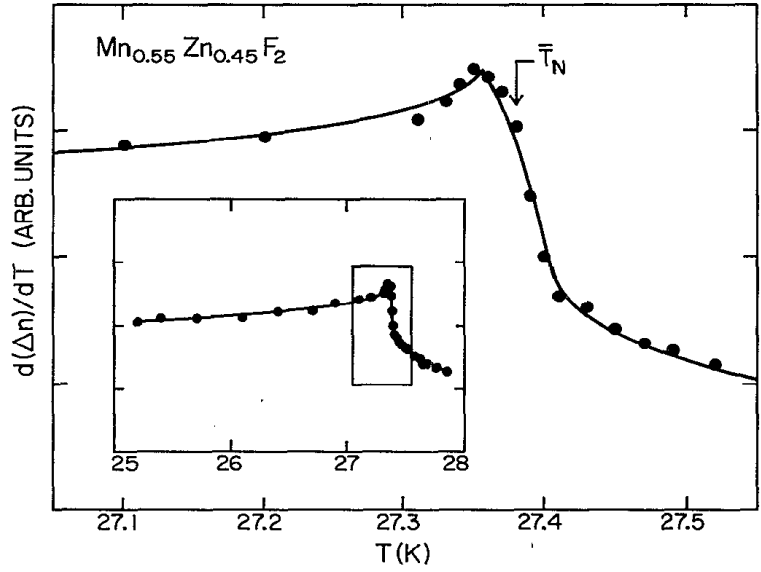}}
\vspace*{5mm}
\caption{\label{figramos}
The temperature derivative of the optical birefringence
of ${\rm Mn_{0.55}Zn_{0.45}F_2}$ at zero magnetic field. In the main part of
the figure ${\rm d} (\Delta n)/{\rm d} T$ is shown for the critical region
(indicated by box in the inset). The measurements bring about a cusp-like
behaviour of the RIM specific heat with the critical exponent
$\alpha= -0.09\pm0.03$. The figure is taken from
Ref. \protect\cite{Ramos88}.
}
\end{figure}

\newpage
\vspace*{10mm}

\newpage
\vspace*{10mm}

\vspace*{10mm}
\def\emmv{\special{em:moveto}}
\def\emln{\special{em:lineto}}
\def\tmls{\footnotesize}
\unitlength=0.7in
\special{em:linewidth 0.4pt}
\linethickness{0.4pt}
\begin{figure}
\begin{center}
\special{em:linewidth 0.4pt}
\linethickness{0.4pt}
\def\emmv{\special{em:moveto}}
\def\emln{\special{em:lineto}}
\def\tmls{\footnotesize}
\begin{picture}(4.9,4.6)(1.1,0.9)
\special{em:linewidth 0.4pt}
\linethickness{0.4pt}
\put(2,1.5){\emmv}
\put(6,1.5){\emln}
\put(2,1.5){\emmv}
\put(2,1.613){\emln}
\put(1.775,1.15){\makebox(0,0)[lb]{\tmls 0.0}}
\put(2.2,1.5){\emmv}
\put(2.2,1.575){\emln}
\put(2.4,1.5){\emmv}
\put(2.4,1.575){\emln}
\put(2.6,1.5){\emmv}
\put(2.6,1.575){\emln}
\put(2.8,1.5){\emmv}
\put(2.8,1.575){\emln}
\put(3,1.5){\emmv}
\put(3,1.613){\emln}
\put(2.775,1.15){\makebox(0,0)[lb]{\tmls 1.0}}
\put(3.2,1.5){\emmv}
\put(3.2,1.575){\emln}
\put(3.4,1.5){\emmv}
\put(3.4,1.575){\emln}
\put(3.6,1.5){\emmv}
\put(3.6,1.575){\emln}
\put(3.8,1.5){\emmv}
\put(3.8,1.575){\emln}
\put(4,1.5){\emmv}
\put(4,1.613){\emln}
\put(3.775,1.15){\makebox(0,0)[lb]{\tmls 2.0}}
\put(4.2,1.5){\emmv}
\put(4.2,1.575){\emln}
\put(4.4,1.5){\emmv}
\put(4.4,1.575){\emln}
\put(4.6,1.5){\emmv}
\put(4.6,1.575){\emln}
\put(4.8,1.5){\emmv}
\put(4.8,1.575){\emln}
\put(5,1.5){\emmv}
\put(5,1.613){\emln}
\put(4.775,1.15){\makebox(0,0)[lb]{\tmls 3.0}}
\put(5.2,1.5){\emmv}
\put(5.2,1.575){\emln}
\put(5.4,1.5){\emmv}
\put(5.4,1.575){\emln}
\put(5.6,1.5){\emmv}
\put(5.6,1.575){\emln}
\put(5.8,1.5){\emmv}
\put(5.8,1.575){\emln}
\put(6,1.5){\emmv}
\put(6,1.613){\emln}
\put(5.775,1.15){\makebox(0,0)[lb]{\tmls 4.0}}
\special{em:linewidth 0.4pt}
\linethickness{0.4pt}
\put(3.936,0.9){\makebox(0,0)[lb]{\tmls d}}
\special{em:linewidth 0.4pt}
\linethickness{0.4pt}
\put(2,1.5){\emmv}
\put(2,5.5){\emln}
\put(2,1.5){\emmv}
\put(2.112,1.5){\emln}
\put(1.2,1.425){\makebox(0,0)[lb]{\tmls 0.00}}
\put(2,1.7){\emmv}
\put(2.075,1.7){\emln}
\put(2,1.9){\emmv}
\put(2.075,1.9){\emln}
\put(2,2.1){\emmv}
\put(2.075,2.1){\emln}
\put(2,2.3){\emmv}
\put(2.075,2.3){\emln}
\put(2,2.5){\emmv}
\put(2.112,2.5){\emln}
\put(1.2,2.425){\makebox(0,0)[lb]{\tmls 0.25}}
\put(2,2.7){\emmv}
\put(2.075,2.7){\emln}
\put(2,2.9){\emmv}
\put(2.075,2.9){\emln}
\put(2,3.1){\emmv}
\put(2.075,3.1){\emln}
\put(2,3.3){\emmv}
\put(2.075,3.3){\emln}
\put(2,3.5){\emmv}
\put(2.112,3.5){\emln}
\put(1.2,3.425){\makebox(0,0)[lb]{\tmls 0.50}}
\put(2,3.7){\emmv}
\put(2.075,3.7){\emln}
\put(2,3.9){\emmv}
\put(2.075,3.9){\emln}
\put(2,4.1){\emmv}
\put(2.075,4.1){\emln}
\put(2,4.3){\emmv}
\put(2.075,4.3){\emln}
\put(2,4.5){\emmv}
\put(2.112,4.5){\emln}
\put(1.2,4.425){\makebox(0,0)[lb]{\tmls 0.75}}
\put(2,4.7){\emmv}
\put(2.075,4.7){\emln}
\put(2,4.9){\emmv}
\put(2.075,4.9){\emln}
\put(2,5.1){\emmv}
\put(2.075,5.1){\emln}
\put(2,5.3){\emmv}
\put(2.075,5.3){\emln}
\put(2,5.5){\emmv}
\put(2.112,5.5){\emln}
\put(1.2,5.425){\makebox(0,0)[lb]{\tmls 1.00}}
\special{em:linewidth 0.4pt}
\linethickness{0.4pt}
\special{em:linewidth 0.4pt}
\linethickness{0.4pt}
\put(2.1,5.454){\emmv}
\put(2.2,5.409){\emln}
\put(2.3,5.364){\emln}
\put(2.4,5.319){\emln}
\put(2.5,5.275){\emln}
\put(2.6,5.23){\emln}
\put(2.7,5.186){\emln}
\put(2.8,5.143){\emln}
\put(2.9,5.099){\emln}
\put(3,5.056){\emln}
\put(3.1,5.012){\emln}
\put(3.2,4.969){\emln}
\put(3.3,4.926){\emln}
\put(3.4,4.883){\emln}
\put(3.5,4.84){\emln}
\put(3.6,4.797){\emln}
\put(3.7,4.755){\emln}
\put(3.8,4.712){\emln}
\put(3.9,4.668){\emln}
\put(4,4.625){\emln}
\put(4.1,4.582){\emln}
\put(4.2,4.538){\emln}
\put(4.3,4.494){\emln}
\put(4.4,4.449){\emln}
\put(4.48,4.413){\emln}
\put(4.5,4.404){\emln}
\put(4.6,4.359){\emln}
\put(4.7,4.312){\emln}
\put(4.8,4.265){\emln}
\put(4.9,4.216){\emln}
\put(5,4.167){\emln}
\put(5.1,4.115){\emln}
\put(5.2,4.062){\emln}
\put(5.3,4.007){\emln}
\put(5.4,3.949){\emln}
\put(5.5,3.888){\emln}
\put(5.6,3.823){\emln}
\put(5.7,3.754){\emln}
\put(5.8,3.678){\emln}
\put(5.9,3.594){\emln}
\special{em:linewidth 0.4pt}
\linethickness{0.4pt}
\put(2.1,2.209){\emmv}
\put(2.2,2.196){\emln}
\put(2.3,2.183){\emln}
\put(2.4,2.17){\emln}
\put(2.5,2.157){\emln}
\put(2.6,2.144){\emln}
\put(2.7,2.131){\emln}
\put(2.8,2.118){\emln}
\put(2.9,2.106){\emln}
\put(3,2.093){\emln}
\put(3.1,2.08){\emln}
\put(3.2,2.067){\emln}
\put(3.3,2.053){\emln}
\put(3.4,2.04){\emln}
\put(3.5,2.027){\emln}
\put(3.6,2.014){\emln}
\put(3.7,2){\emln}
\put(3.8,1.986){\emln}
\put(3.9,1.973){\emln}
\put(4,1.959){\emln}
\put(4.1,1.944){\emln}
\put(4.2,1.93){\emln}
\put(4.3,1.915){\emln}
\put(4.4,1.899){\emln}
\put(4.48,1.887){\emln}
\put(4.5,1.884){\emln}
\put(4.6,1.868){\emln}
\put(4.7,1.851){\emln}
\put(4.8,1.833){\emln}
\put(4.9,1.815){\emln}
\put(5,1.796){\emln}
\put(5.1,1.776){\emln}
\put(5.2,1.755){\emln}
\put(5.3,1.733){\emln}
\put(5.4,1.708){\emln}
\put(5.5,1.682){\emln}
\put(5.6,1.653){\emln}
\put(5.7,1.622){\emln}
\put(5.8,1.586){\emln}
\put(5.9,1.546){\emln}
\special{em:linewidth 0.4pt}
\linethickness{0.4pt}
\put(4,2.1){\makebox(0,0)[lb]{\tmls $-i_2$}}
\special{em:linewidth 0.4pt}
\linethickness{0.4pt}
\put(4,4.3){\makebox(0,0)[lb]{\tmls $i_1$}}
\end{picture}
\end{center}
\vspace*{5mm}
\caption[]{\label{figloopint} Two--loop integrals of the RG
functions in massive scheme (formulas
(\protect\ref{43})--(\protect\ref{46})) as functions of space
dimension $d$, Ref. \protect\cite{Holovatch92}. }
\end{figure}
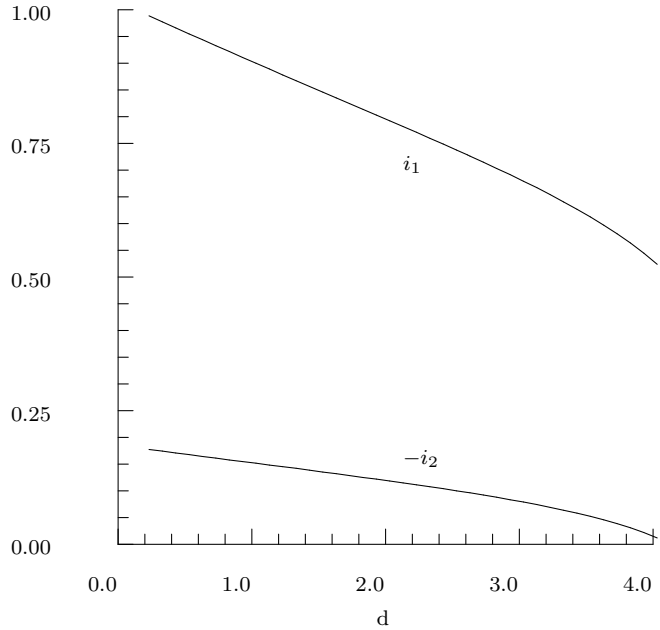

\newpage
\vspace*{10mm}
\begin{figure}[ht]
\begin{centering}
\setlength{\unitlength}{1mm}
\begin{picture}(126,110)
\epsfxsize=126mm
\put(10,-10){\epsffile[22 20 822 582]{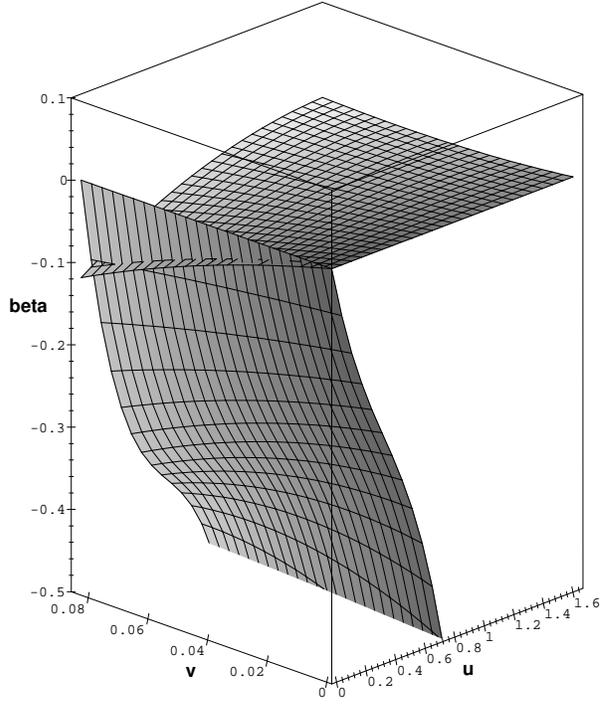}}
\end{picture}\\
\end{centering}
\vspace*{5mm}
\caption{\label{figbetanon} $\beta$-functions of RIM
$\beta_u(u,v)$, $\beta_v(u,v)$ calculated in two--loop
approximation by a $d=3$ minimal subtraction scheme without
resummation. Only the
Gaussian fixed point $u^*=v^*=0$ survives. The figure is taken
from Ref. \protect\cite{Folk99a}. }
\end{figure}

\newpage
\vspace*{10mm}

\vspace*{15mm}

\newpage
\vspace*{10mm}
\begin{figure}[ht]
\begin{centering}
\setlength{\unitlength}{1mm}
\begin{picture}(126,110)
\epsfxsize=126mm
\put(10,0){\epsffile[22 20 822 582]{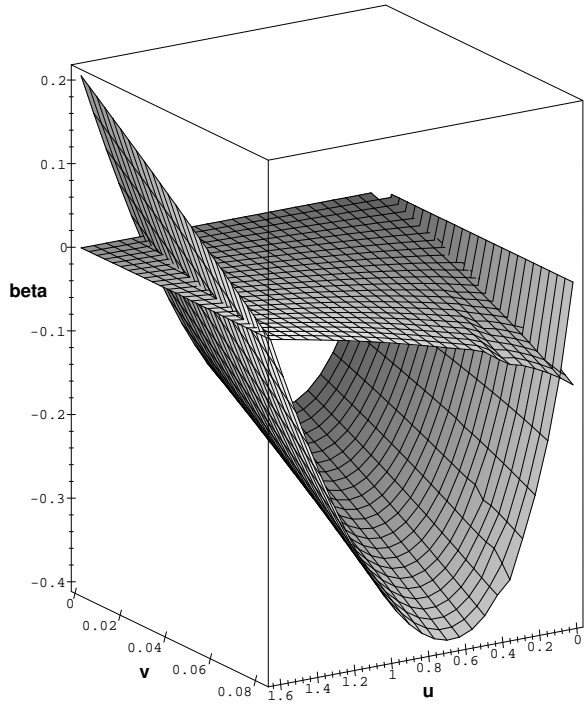}}
\end{picture}\\
\end{centering}
\vspace*{5mm}
\caption{\label{figbetares} Pad\'e-Borel resummation technique
applied to the $\beta$-functions of the figure \ref{figbetanon}.
Resummation restores the presence of the fixed point $u^*\neq 0,
\, v^*=0$ and results in the appearance of a new stable fixed
point  $u^*\neq 0, \, v^*\neq 0$. The figure is taken from
Ref. \protect\cite{Folk99a}.
Note that the authors of Ref. \protect\cite{Folk99a} exploited different
in comparison to formulas
(\protect\ref{35})-(\protect\ref{38})
normalization of couplings $u\rightarrow 3/2\,u$ and $v\rightarrow-4\,v$.
}
\end{figure}

\newpage
\vspace*{10mm}

\vspace*{15mm}

\newpage
\vspace*{10mm}

\newpage
\vspace*{20mm}
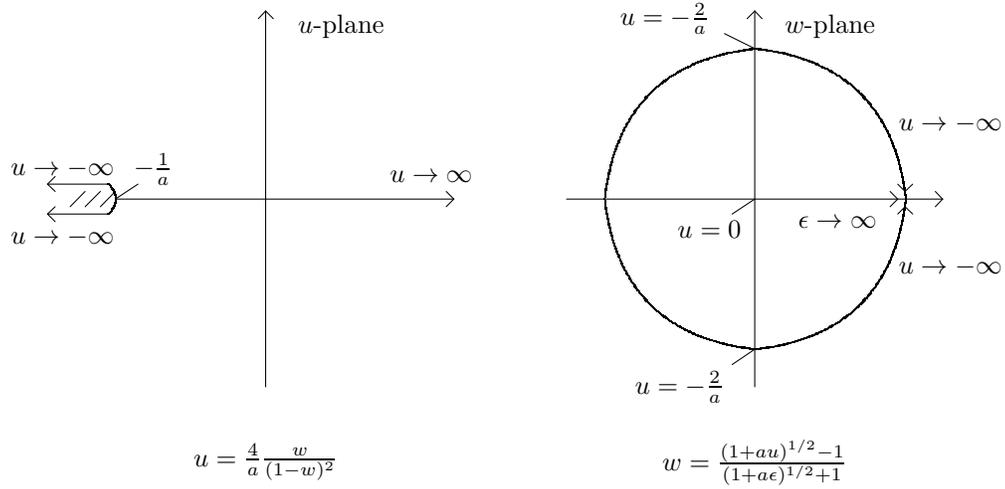
\begin{figure} [htbp]
\begin{center}
\unitlength=1.00mm
\special{em:linewidth 0.4pt}
\linethickness{0.4pt}
\begin{picture}(122.00,105.00)
\emline{71.00}{80.00}{1}{121.00}{80.00}{2}
\emline{96.00}{55.00}{3}{96.00}{105.00}{4}
\bezier{144}(76.00,80.00)(78.00,98.00)(96.00,100.00)
\bezier{144}(96.00,100.00)(114.00,98.00)(116.00,80.00)
\bezier{144}(116.00,80.00)(114.00,62.00)(96.00,60.00)
\bezier{144}(96.00,60.00)(78.00,62.00)(76.00,80.00)
\emline{56.00}{80.00}{5}{11.00}{80.00}{6}
\emline{11.00}{81.00}{7}{9.00}{79.00}{8}
\emline{9.00}{81.00}{9}{7.00}{79.00}{10}
\emline{7.00}{81.00}{11}{5.00}{79.00}{12}
\bezier{24}(10.00,82.00)(12.00,80.00)(10.00,78.00)
\emline{115.00}{80.00}{13}{114.00}{81.00}{14}
\emline{115.00}{80.00}{15}{114.00}{79.00}{16}
\emline{116.00}{81.00}{17}{115.00}{82.00}{18}
\emline{116.00}{81.00}{19}{117.00}{82.00}{20}
\emline{116.00}{79.00}{21}{115.00}{78.00}{22}
\emline{116.00}{79.00}{23}{117.00}{78.00}{24}
\emline{121.00}{80.00}{25}{120.00}{81.00}{26}
\emline{121.00}{80.00}{27}{120.00}{79.00}{28}
\emline{96.00}{105.00}{29}{95.00}{104.00}{30}
\emline{96.00}{105.00}{31}{97.00}{104.00}{32}
\emline{31.00}{55.00}{33}{31.00}{105.00}{34}
\emline{31.00}{105.00}{35}{30.00}{104.00}{36}
\emline{31.00}{105.00}{37}{32.00}{104.00}{38}
\emline{56.00}{80.00}{39}{55.00}{81.00}{40}
\emline{56.00}{80.00}{41}{55.00}{79.00}{42}
\emline{11.00}{81.00}{43}{11.00}{79.00}{44}
\emline{10.00}{82.00}{45}{2.00}{82.00}{46}
\emline{2.00}{82.00}{47}{3.00}{83.00}{48}
\emline{2.00}{82.00}{49}{3.00}{81.00}{50}
\emline{10.00}{78.00}{51}{2.00}{78.00}{52}
\emline{2.00}{78.00}{53}{3.00}{77.00}{54}
\emline{2.00}{78.00}{55}{3.00}{79.00}{56}
\put(107.00,77.00){\makebox(0,0)[cc]{$\epsilon \rightarrow \infty$}}
\put(84.00,104.00){\makebox(0,0)[cc]{$u=-\frac{2}{a}$}}
\emline{96.00}{100.00}{57}{92.00}{102.00}{58}
\emline{96.00}{80.00}{59}{93.00}{78.00}{60}
\put(90.00,76.00){\makebox(0,0)[cc]{$u = 0$}}
\put(122.00,71.00){\makebox(0,0)[cc]{$u \rightarrow -\infty$}}
\put(122.00,90.00){\makebox(0,0)[cc]{$u \rightarrow -\infty$}}
\emline{96.00}{60.00}{61}{93.00}{58.00}{62}
\put(86.00,55.00){\makebox(0,0)[cc]{$u=-\frac{2}{a}$}}
\put(53.00,83.00){\makebox(0,0)[cc]{$u \rightarrow \infty$}}
\put(4.00,84.00){\makebox(0,0)[cc]{$u \rightarrow -\infty$}}
\put(4.00,75.00){\makebox(0,0)[cc]{$u \rightarrow -\infty$}}
\put(106.00,103.00){\makebox(0,0)[cc]{$w$-plane}}
\put(41.00,103.00){\makebox(0,0)[cc]{$u$-plane}}
\emline{11.00}{80.00}{63}{14.00}{82.00}{64}
\put(16.00,84.00){\makebox(0,0)[cc]{$-\frac{1}{a}$}}
\put(31.00,45.00){\makebox(0,0)[cc]{$u=\frac{4}{a}\frac{w}{(1-w)^2}$}}
\put(96.00,45.00){\makebox(0,0)[cc]
{$w=\frac{(1+au)^{1/2} - 1}{(1+a\epsilon)^{1/2}+1}$}}
\end{picture}
\end{center}
\vspace*{-10mm}
\caption {\label{figconfmap} Conformal mapping of the cut-plane
onto a disc leaving the origin invariant, as prescribed by
formulas (\ref{65}). See the text for details. The figure is taken
from Ref. \protect\cite{Holovatch96}. }
\end{figure}

\begin{thebibliography}{199}
\bibitem{note1}
We give here recent renormalization group estimates for the
critical exponents of the pure three dimensional Ising model obtained from
the resummation of perturbation theory series of a scalar $d=3$ $\phi^4$ theory
\cite{Guida98}.
\bibitem{note2}
Currently two possible scenarios for the influence of long-range-correlated
disorder on critical behaviour are discussed (c.f. Refs \cite{Weinrib83} and
\cite{longmass}). Although  both confirm the relevance of
long-range-correlated disorder, they lead to different numerical predictions
for the critical exponents.
\end{thebibliography}
\end{document}